\documentclass{aa} 
 
\usepackage[varg]{txfonts} 
\usepackage{natbib} 
\usepackage{graphicx}

\bibpunct{(}{)}{;}{a}{}{,}

\newcommand{\unit}{\mathrm} 
\newcommand{\name}{\mathrm} 
\newcommand{\dd}{\mathrm{d}}

\begin{document}

\title{Substellar fragmentation in self-gravitating fluids with a major phase transition} 
\author{A. F\"{u}glistaler \and D. Pfenniger} 
\institute{Geneva Observatory, University of Geneva, Sauverny, Switzerland} 
\date{Received 12 August 2014 / Accepted 22 March 2015}

\abstract 
{The observation of various ices in cold molecular clouds, the existence of ubiquitous substellar, cold H$_2$ globules
  in planetary nebulae and supernova remnants, or the mere existence of comets suggest that the physics of very cold
  interstellar gas might be much richer than usually envisioned.  At the extreme of low temperatures
  ($\lesssim10\,\unit{K}$), H$_2$ itself is subject to a phase transition crossing the entire cosmic gas density
  scale. }
{This well-known, laboratory-based fact motivates us to study the ideal case of a cold neutral gaseous medium in
  interstellar conditions for which the bulk of the mass, instead of trace elements, is subject to a gas-liquid or
  gas-solid phase transition.}
{On the one hand, the equilibrium of general non-ideal fluids is studied using the virial theorem and linear stability
  analysis.  On the other hand, the non-linear dynamics is studied using computer simulations to characterize the
  expected formation of solid bodies analogous to comets. The simulations are run with a state-of-the-art molecular
  dynamics code (LAMMPS) using the Lennard-Jones inter-molecular potential.  The long-range gravitational forces can be
  taken into account together with short-range molecular forces with finite limited computational resources, using
  super-molecules, provided the right scaling is followed.}
{The concept of super-molecule, where the phase transition conditions are preserved by the proper choice of the particle
  parameters, is tested with computer simulations, allowing us to correctly satisfy the Jeans instability criterion for
  one-phase fluids.  The simulations show that fluids presenting a phase transition are gravitationally unstable as
  well, independent of the strength of the gravitational potential, producing two distinct kinds of substellar bodies,
  those dominated by gravity (``planetoids'') and those dominated by molecular attractive force (``comets'').}
{Observations, formal analysis, and computer simulations suggest the possibility of the formation of substellar H$_2$
  clumps in cold molecular clouds due to the combination of phase transition and gravity.  Fluids presenting a phase
  transition are gravitationally unstable, independent of the strength of the gravitational potential. Arbitrarily small
  H$_2$ clumps may form even at relatively high temperatures up to $400$ -- $600\,\unit{K}$, according to virial
  analysis. The combination of phase transition and gravity may be relevant for a wider range of astrophysical
  situations, such as proto-planetary disks.}

\keywords{Instabilities -- ISM: clouds -- ISM: kinematics and dynamics -- ISM: molecules -- Methods: analytical --
  Methods: numerical}

\maketitle

\section{Introduction} 

Typically around $50\%$ or more of the gas in spiral galaxies consists of H$_2$, inferred indirectly by CO or dust
emission. Since the discovery of dark molecular hydrogen \citep{grenier_unveiling_2005, langer_c+_2010,
  planck_collaboration_planck_2011, paradis_dark_2012}, the estimated quantity of H$_2$ in the Milky Way has essentially
been doubled, effectively revealing the nature of some of the dark baryons.  It is commonly admitted that the CO-related
molecular hydrogen is present in relatively dense regions of the interstellar medium, molecular clouds with number
density $> 10^{10}\,\unit{m^{-3}}$ and temperatures of $7$--$30\,\unit{K}$ \citep{draine_physics_2011}.

Even though H$_2$ is by far the most abundant molecule ($\sim 90\%$), molecular clouds are mainly detected by CO
emissions because of all the difficulties in detecting cold H$_2$ \citep{bolatto_co--h_2013}. For example, H$_2$ only
starts emitting at temperatures $> 512\,\unit{K}$. See \cite{combes_perspectives_1997} for a review of several possible
methods for detecting cold H$_2$.  Because of these detection difficulties, the real quantity of H$_2$ (as well as He,
which shares similar properties of discreteness with H$_2$) in molecular clouds is still rather unknown, especially when
the gas temperature is below $\lesssim 8\,\unit{K}$ down to the cosmic background temperature of $2.76\,\unit{K}$.

\begin{figure}[t] 
\resizebox{\hsize}{!}{\includegraphics{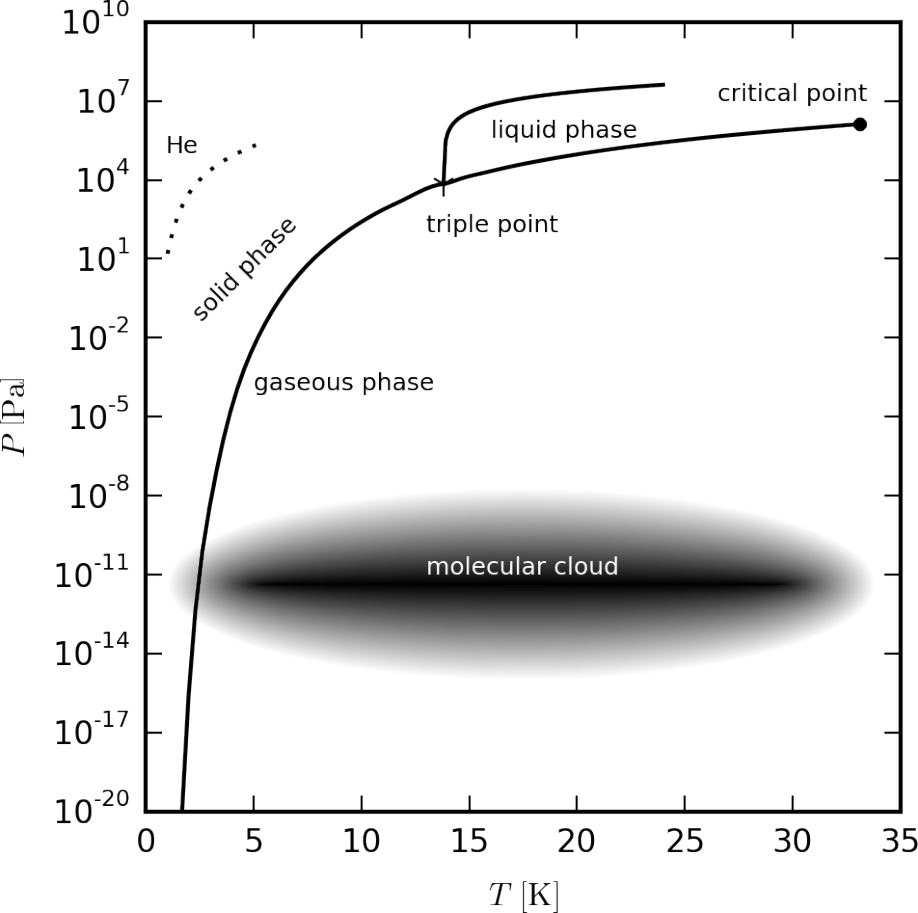}} 
\caption{H$_2$ phase diagram (bold line) and He (dotted line) in cold and low pressure conditions. }
\label{fcc} 
\end{figure}

The condensation properties of H$_2$ relevant for molecular cloud conditions are well known from laboratory data
\citep{air_liquide_gas_1976}.  The phase diagram (Fig.~\ref{fcc}) shows the domain of pressure conventionally attributed
to molecular clouds.  One has to keep in mind, however, that since molecular clouds are highly structured
\citep{pfenniger_is_1994}, and commonly observed in a state of supersonic turbulence
\citep{elmegreen_interstellar_2004}, large fluctuations in density and temperature must occur.

\begin{figure}[t] 
\resizebox{\hsize}{!}{\includegraphics{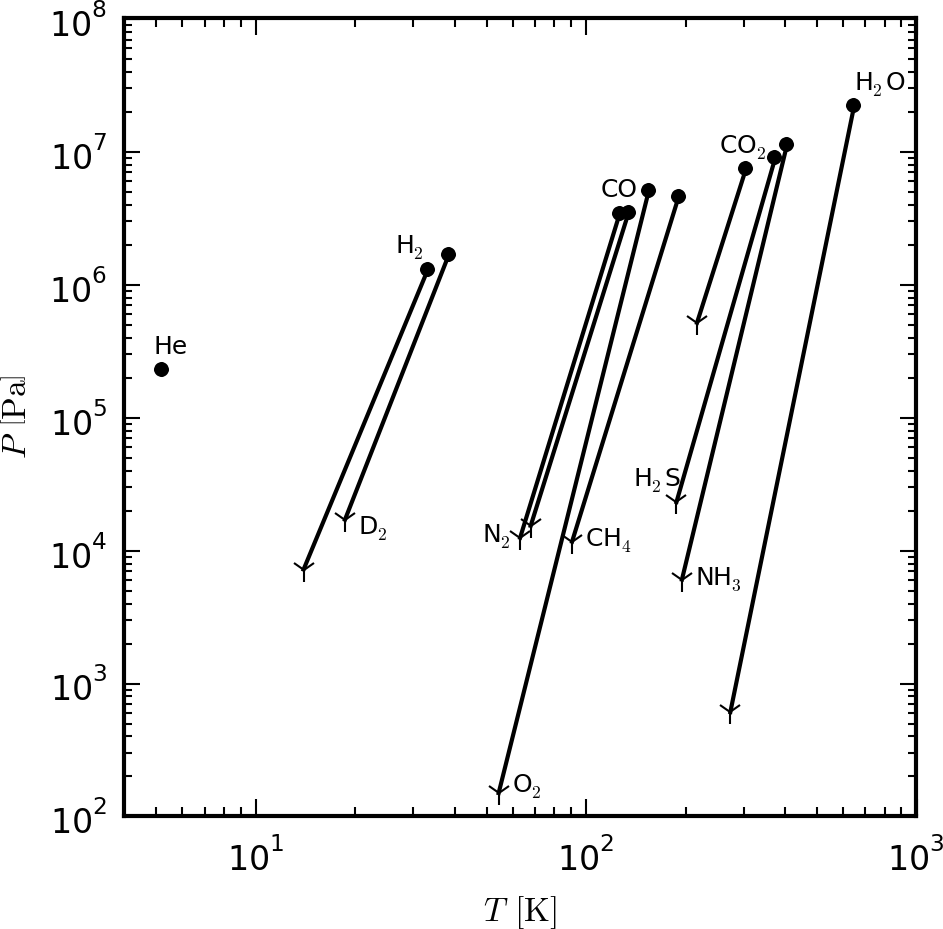}} 
\caption{Critical points (bullet) and triple points ($\scriptstyle{\mathsf{Y}}$) of common molecules in the ISM. As for
  H$_2$ in Fig.~\ref{fcc}, the sublimation curves cross interstellar pressure conditions very steeply in pressure a few
  K below their triple point.  For example, CO should essentially be frozen below $\sim 20\,\unit{K}$, so unable to emit
  rotation lines.}
\label{fmol}
\end{figure}

The presence of ice in the interstellar medium consisting of heavier molecules, such as H$_2$O, CO, CO$_2$ and NH$_3$
covering dust grains, is nowadays well documented \citep{allamandola_evolution_1999}. Figure \ref{fmol} shows the
location of the critical and triple points of abundant molecules existing in the ISM. H$_2$ ice has been detected in the
absorption band at $2.417\,\mu\unit{m}$ \citep{sandford_h2_1993, buch_interpretation_1994, dissly_h2-rich_1994}, but the
interpretation of this detection is that H$_2$ is mixed within H$_2$O-rich grains in conditions that are too warm to
allow the bulk of H$_2$ to condense \citep{kristensen_h_2011}.

High-resolution pictures of nearby planetary nebulae or remnants of supernova have shown the presence of substellar
fragments \citep{walsh_imaging_1993}. These very cold globules, or knots, each of the size of a few tens of AU, are at
least as cold in the inner parts ($\sim 10\,\unit{K}$) as molecular clouds, but much smaller.  An important feature is
that the apparent column density in these knots increases inwards as long as the resolution allows, or until the knot is
optically thin \citep{burkert_structure_1998}.  If these trends extend to the centre, one can expect there would be much
higher density at colder conditions. It would be ideal to eventually reach a regime where H$_2$ could condense in liquid
or solid form, especially because at high column density the medium blocks UV radiation and cosmic ray heating.

At the level of molecular clouds, even though their average properties are well separated from the H$_2$ phase
transition, these are only static properties that ignore the highly dynamical nature of supersonic turbulence observed
in the interstellar medium, where fluctuations must be large.  Lower temperatures can be reached with fast adiabatic
decompression alone.  An example of fast decompression is displayed in the Boomerang Nebula, which reaches a temperature
of only $1\,\unit{K}$ because of a fast expanding wind \citep{sahai_boomerang_1997}. Thus, one should expect, to be
coherent with supersonic turbulence, that regions of expansion and compression must be common in the ISM.

The fragmentation of self-gravitating gas leading to collapse has been studied for over a century. The molecular cloud
is normally represented as a self-gravitating, one-phase fluid (i.e. pure gas), which is governed by the balance between
gravity and gas pressure. The gravity of this fluid is directly proportional to its density, and the pressure to the
adiabatic density derivative of the pressure, $(\partial P/ \partial\rho)_\name{s}$. It has been shown by
\citet{jeans_stability_1902} that if a perturbation with a long enough wavelength is introduced into the system, this
perturbation will grow exponentially. With growing perturbation, the pressure of the fluid will not be able to withstand
its gravity anymore, which will ultimately lead to gravitational collapse.

The approximation of molecular clouds as a one-phase fluid is valid in many cases, but when considering very cold,
high-density regions, it is an over-simplification. In these condition, H$_2$ will start creating ice, which has to be
taken into account \citep{walker_snowflakes_2013}.  The dynamics of this kind of fluid presenting a phase transition is
different from a one-phase fluid, the most important difference being a very low value of
$(\partial P/\partial\rho)_\name{s}$ \citep{johnston_thermodynamic_2014}. But, as $(\partial P/\partial\rho)_\name{s}$
is crucial for the stability of a fluid, a cold high-density fluid presenting a phase transition is therefore expected
to be very unstable.

Based on the observation of substellar globules in planetary nebulas and the dynamics of fluids presenting a phase
transition, we may expect the presence of small, substellar H$_2$ ice fragments in molecular clouds due to the
fragmentation of cold high-density regions. It is of great astronomical interest to study the nature of this substellar
fragmentation, as the resulting bound objects may be too small to start nuclear fusion and, thanks to their very low
temperature, are very difficult to detect. Some of the baryonic dark matter may actually consist of these fragments
\citep{pfenniger_is_1994, pfenniger_is_1994-1}.

In this article, we study the physics of a self-gravitating van der Waals fluid presenting a phase transition
analytically and with simulations. In the analytic part (Sect.~\ref{sPT}), the physics of a van der Waals fluid and the
related Lennard-Jones (thereafter LJ) potential are recalled. We calculate the virial theorem taking both the
gravitational and the LJ potential into account. The virial analysis helps to characterize different types of
fluids. The stability of a self-gravitating van der Waals fluid presenting a phase transition is then analyzed.

In Sect.~\ref{sMD}, the molecular dynamics simulator LAMMPS \citep{plimpton_fast_1995} is introduced. By proper scaling
of physical constants the Coulomb force solver is used to calculate the gravitational force, and the short rang
molecular force is calculated with the LJ force. We introduce the concept of super-molecules to enable us to perform
simulations with a total mass high enough for the fluid to be self-gravitating.

The simulations performed are discussed in Sect.~\ref{sS}. First, the correctness of the super-molecule approach is
tested. Second, one-phase fluids (i.e.\ pure gas) are used to test the Jeans criterion. Third, simulations close to a
phase transitions are performed, studying the properties of non-gravitating fluids and fluids with a gravitational
potential above and below the Jeans criterion.

\section{Physics of a fluid presenting a phase transition} 
\label{sPT} 

\begin{figure}[t] 
  \resizebox{\hsize}{!}{\includegraphics{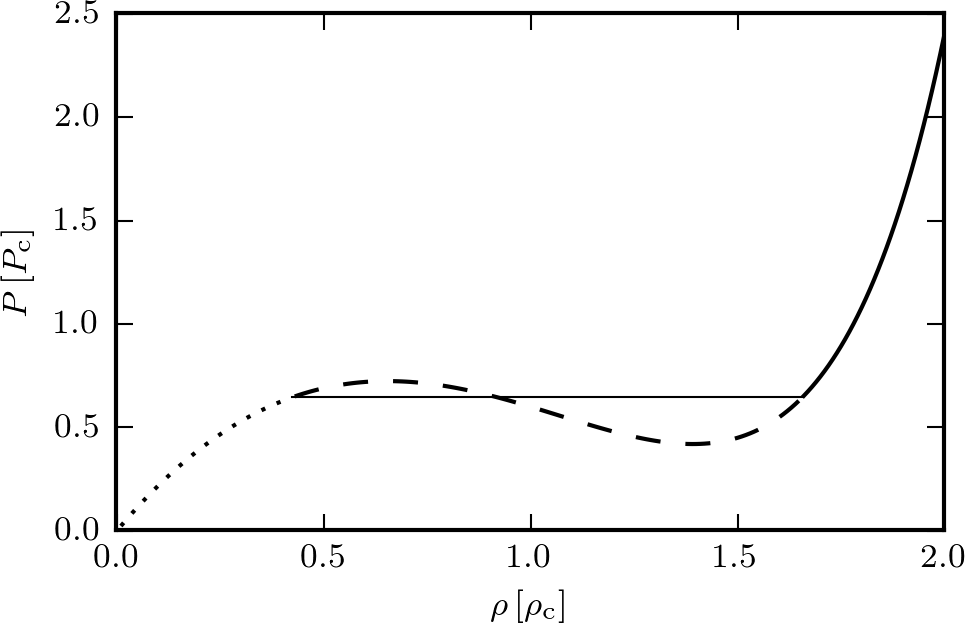}}
  \caption{van der Waals phase diagram for a fluid with $T_\name{r}=0.9$. Dotted line: gas phase; dashed line: phase
    transition; solid line: solid phase.}
	\label{fvdw2d}
\end{figure}

\begin{figure}[t] 
  \resizebox{\hsize}{!}{\includegraphics{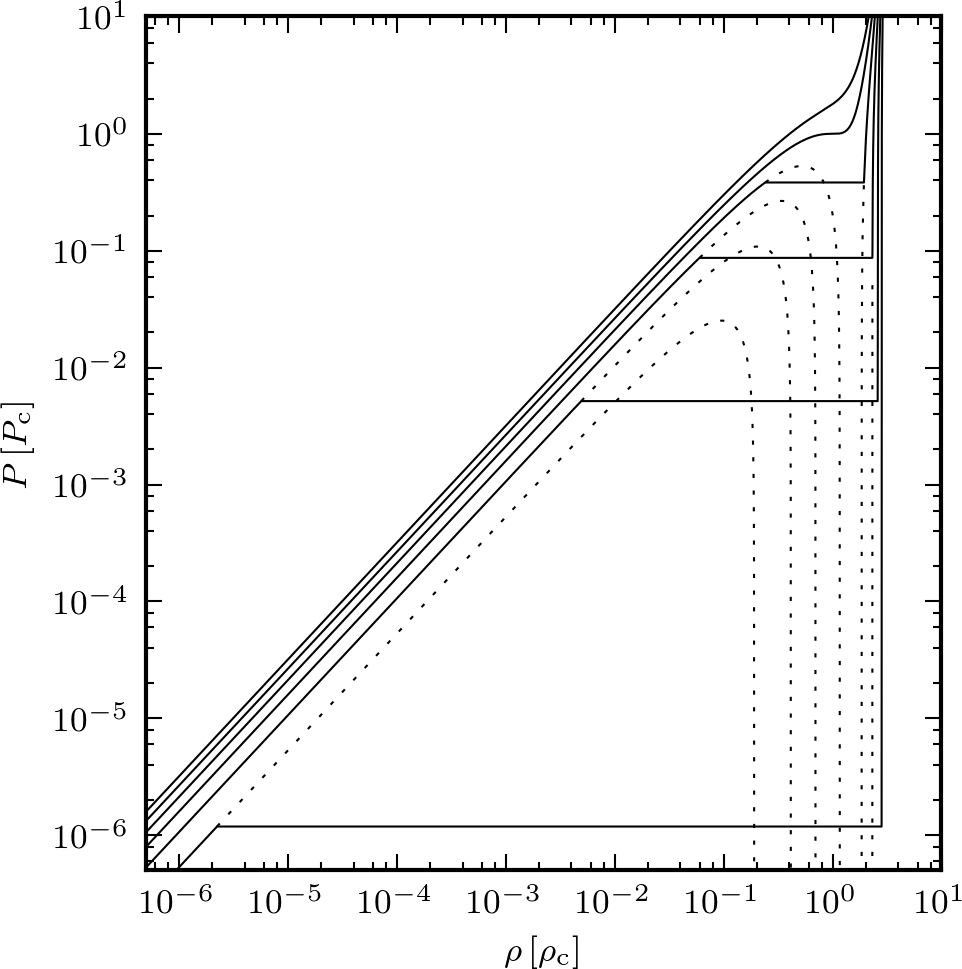}}
  \caption{van der Waals phase diagram for a fluid with a temperature (from bottom to top) of $T_\name{r}=0.2$,
    $T_\name{r}=0.4$, $T_\name{r}=0.6$, $T_\name{r}=0.8$, $T_\name{r}=1.0$, and $T_\name{r}=1.2$. Solid line: van der
    Waals EOS with Maxwell construct, dotted line: original van der Waals EOS.}
        \label{fvdw}
\end{figure}

\begin{figure}[t] 
  \resizebox{\hsize}{!}{\includegraphics{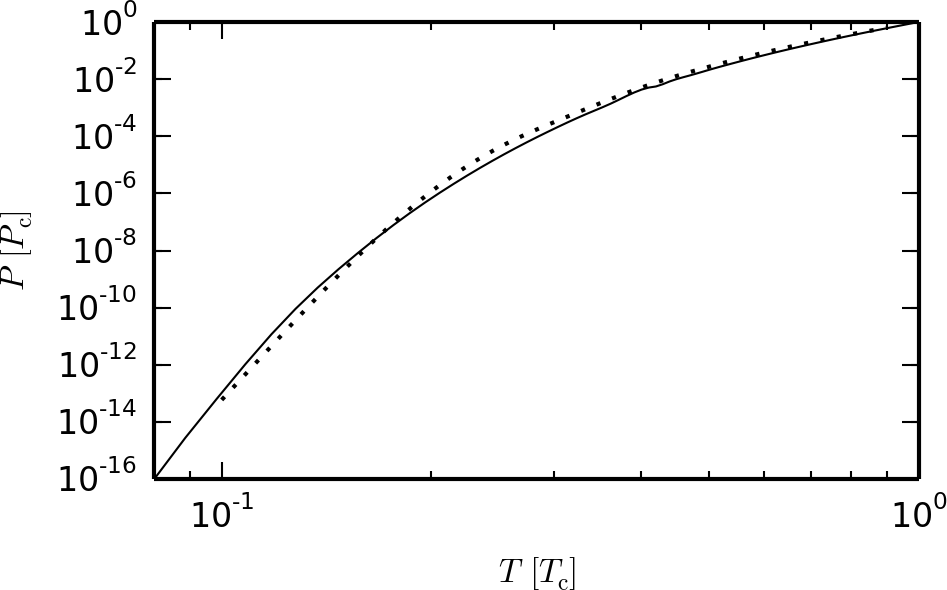}}
  \caption{H$_2$ and van der Waals phase diagrams. Solid line: H$_2$ laboratory data; dotted line: van der Waals vapour
    curve derived with the Maxwell construct.}
        \label{fmaxlab}
\end{figure}

In this section, we describe the physics of a self-gravitating fluid presenting a phase transition.
\subsection{van der Waals equation} 
\label{ssVdW}
The van der Waals equation is a classical equation of state (EOS)\ of a pure fluid presenting a first order phase
transition.  This EOS describes the macroscopic behaviour of a fluid in which at microscopic level the molecules are
strongly repulsive at short distances and beyond that weakly attractive over a limited range.

The van der Waals equation \citep{van_der_waals_remarks_1910,johnston_thermodynamic_2014} is a modification of the
ideal-gas law, taking the finite size of molecules and intermolecular interactions into account. It links pressure,
density, and temperature as follows:
\begin{equation} 
  P = \frac{k_\name{B}T\,n}{1 - bn}-an^2 \label{evdw} \ ,
\end{equation} 
with $a\,[\unit{Pa\,m^6}]$ and $b\,[\unit{m^3}]$ being constants characteristic of the fluid. It can also be expressed
in a reduced, dimensionless form as,
\begin{equation} 
  P_\name{r} = \frac{8T_\name{r}}{\frac{3}{n_\name{r}}-1} - 3n_\name{r}^2 \ ,
\end{equation} 
where $P_\name{r} = P / P_\name{c}$, $n_\name{r} = n / n_\name{c}$, and $T_\name{r} = T / T_\name{c}$. The parameters
$P_\name{c}$, $T_\name{c}$ and $n_\name{c}$ are the values of the thermodynamic critical point
\citep{kondepudi_modern_1998, carey_statistical_1999}.

Figure \ref{fvdw2d} shows the phase diagram for a fluid with $T = 0.9T_\name{c}$. One can distinguish the gaseous phase
(dotted line) and the solid phase (solid line); in between (dashed line), the fluid presents a phase transition. There
are states where $(\partial P/\partial \rho)_\name{s} < 0$, which is thermodynamically unstable. The parts of the dashed
curve where $(\partial P/\partial \rho)_\name{s} > 0$ are metastable because a lower entropy state is reached when the
fluid splits into condensed and gaseous components.  The fraction of condensed over gaseous phase grows from 0 to 1 from
left to right.

When the two phases coexist, the van der Waals EOS is replaced by a constant pressure marked by a horizontal line.  This
constant pressure level is determined by the Maxwell ``equal area'' construct \citep{clerk-maxwell_dynamical_1875},
demanding a total zero $P\cdot v$ work for an adiabatic cycle between the fully gaseous to the fully condensed state.

Figure \ref{fvdw} shows the phase diagrams for fluids with a temperature from $T = 0.2 T_\name{c}$ to
$T = 1.2 T_\name{c}$. The dotted line shows the original van der Waals EOS whereas the solid line displays the modified
law using the Maxwell construct. There is no Maxwell construct for $T \geq T_\name{c}$ as
$(\partial P/\partial \rho)_\name{s}$ is always $\geq 0$. \cite{lekner_parametric_1982} provides a parametric solution
to the van der Waals and Maxwell construct, using $\Delta s$ as variable.

The phase equilibrium pressure is almost identical with the coexistence curve of laboratory data for H$_2$ over a wide
range of pressure (Fig.~\ref{fmaxlab}).

\subsubsection{Lennard-Jones forces} 
\label{ssLJ} 

\begin{figure}[th]
  \resizebox{\hsize}{!}{\includegraphics{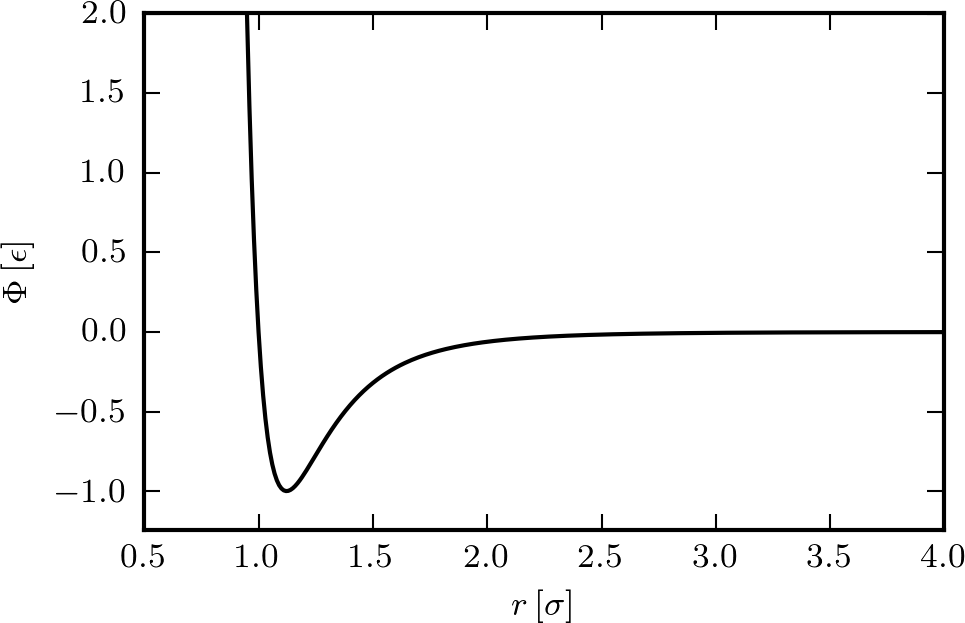}}
  \caption{Lennard-Jones potential.}
  \label{fLJ}
\end{figure}

The van der Waals model of phase transition fluid is convenient for a continuum fluid description, but fails to
represent all the phenomena around the phase transition, which is the reason for correcting it with the above
mentioned Maxwell construct.  Even with the Maxwell construct, however, a local thermal equilibrium is still supposed to
hold.  In astrophysical contexts, often even these assumptions cannot be granted.  For instance, in supersonic turbulent
situations, ubiquitous in the interstellar medium, thermal equilibrium cannot be satisfied since mechanical energy
propagates faster than thermal energy through pressure.  Thus any method using quantities, like temperature or pressure,
implicitly assumes that a local thermal equilibrium is established, which makes their use in the supersonic turbulent
regime uncertain.

A much less demanding model of fluid is provided by molecular dynamics, where the simplest molecule interactions close
to the van der Waals model in equilibrium situations is provided by the Lennard-Jones (LJ) intermolecular potential
\citep{jones_determination_1924}.  No local or global thermal equilibirum is required.  The LJ potential consists of an
attractive long-range term and a repulsive short-range term (Fig.~\ref{fLJ}),
\begin{equation} 
  \Phi_{\name{LJ}}(r) = 4\frac{\epsilon}{m}\left(r_\sigma^{-12} - r_\sigma^{-6}\right) \ ,
\end{equation} 
with $r_\sigma = r/\sigma$. Its minimum value, located at $r_\sigma = 2^{1/6}$, is $-\epsilon/m$.

\subsection{Virial theorem for a Lennard-Jones gravitating fluid} 
\label{ssV} 

\subsubsection{Lagrange-Jacobi identity}

The virial theorem \citep{clausius_ueber_1870} describes the statistical equilibrium of a system of interacting
particles or fluid systems, relating the kinetic and potential energies.  In the case of a self-gravitating LJ fluid,
the LJ potential and gravity combine as a total potential $\Phi = \Phi_\name{G} +\Phi_\name{LJ}$.

The virial theorem for this new potential can be derived using the Lagrange-Jacobi identity path, by taking the second
time derivative of the moment of the polar inertia $I \equiv \sum_i m_i\vec{r}_i^2$, we find,
\begin{equation} 
  \frac{1}{2} {\dd^2I \over \dd t^2} = \sum_i m_i\vec{\dot{r}}_i^2 + \sum_i m_i\vec{r}_i\cdot \vec{\ddot{r}}_i \ ,
\end{equation}
where the first right-hand term is twice the kinetic energy $E_\name{kin}$, and the second one is the virial term.  For
a system near a statistical equilibrium, both sides of this equation should oscillate around 0, so the respective
time averages should vanish.

The LJ potential is a sum of two homogeneous functions
\footnote{By definition, a homogeneous function $f$ of degree $k$ satisfies
  $f(\lambda \vec{x}) = \lambda^k f(\vec{x})$.}
$\Phi_\name{LJ}=\Phi_\name{LJ, a}+\Phi_\name{LJ, r}$ of degree $-6$ and $-12$ respectively, while gravity is of degree
$-1$.  So we can use Euler's theorem of homogeneous functions to express the virial term as a sum of potential energies
multiplied by minus the homogeneous degree.  The Lagrange-Jacobi identity becomes
\begin{equation} 
  \label{eVir}
  {1 \over 2} {\dd^2I \over \dd t^2} = \underbrace{2 E_\name{kin}}_{>0} + \underbrace{12 E_\name{pot, LJ, r}}_{>0} +
  \underbrace{6 E_\name{pot, LJ, a}}_{<0} + \underbrace{E_\name{pot, G}}_{<0} \ .
\end{equation}

\subsubsection{Homogeneous sphere}

\begin{figure}[t]
  \resizebox{\hsize}{!}{\includegraphics{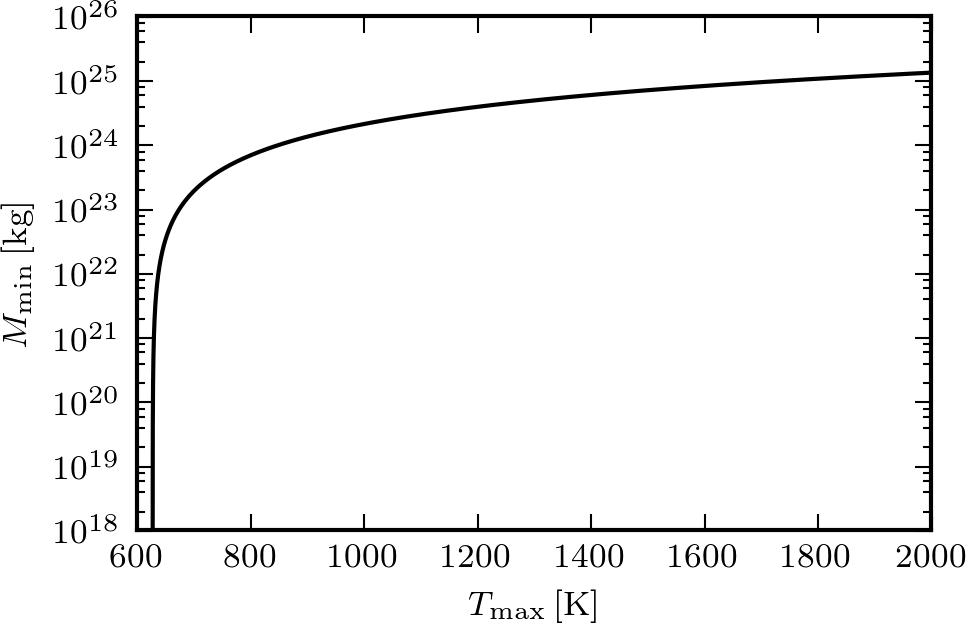}}
  \caption{Minimum mass of isothermal equilibrium curves for H$_2$ homogeneous spheres.}
\label{fMT}
\end{figure}

\begin{figure}[t]
  \resizebox{\hsize}{!}{\includegraphics{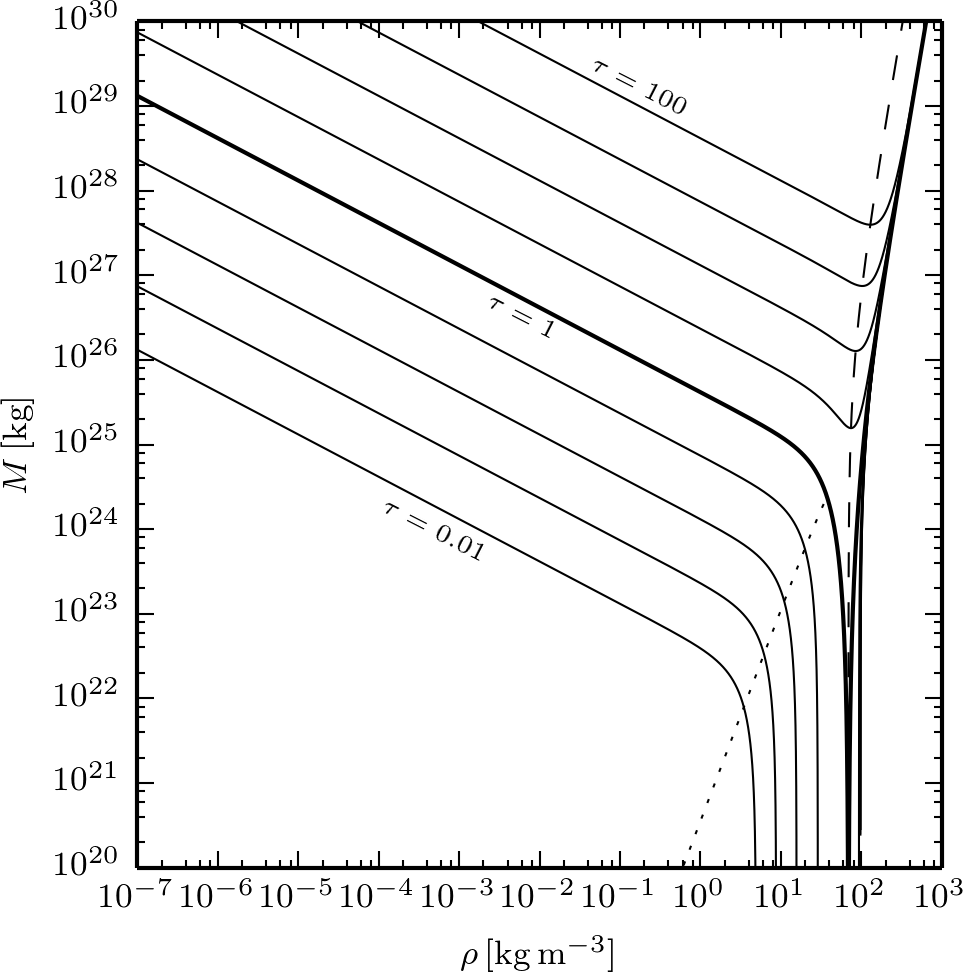}}
  \caption{Virial equilibrium of gravitating LJ homogeneous isothermal spheres with H$_2$ molecules arranged on a
    HCP/FCC lattice. The sphere mass as a function of density is plotted for temperatures ranging from
    $\tau \equiv T/T_\name{max} = 10^{-2}$ to $10^2$.}
\label{fvirial-ratios}
\end{figure}

For homogeneous, finite mass spheres at a given temperature, the individual terms of the Lagrange-Jacobi identity read,
\begin{eqnarray}
  2E_\name{kin} &=&  {3 k_\name{B}T\over m} M \ ,\\
  12 E_\name{pot, LJ, r} &=&  12 c_\name{r} \,  \,{\epsilon \sigma^{12} \rho^4\over m^5} \,M  \ , \label{eELJr}\\
  6 E_\name{pot, LJ, a}  &=& -6 c_\name{a} \,\, {\epsilon \sigma^6 \rho^2\over m^3} \,M  \ , \label{eELJa}\\
  E_\name{pot,G} &=&  -G \left({36 \pi \rho \over 125} \right)^{1/3}  M^{5/3}  \ , \label{eEgrav}
\end{eqnarray}
where the constants $c_\name{r} $ and $c_\name{a}$ in the LJ terms have been calculated by straight summation over the
$1.5\cdot 10^{12}$ nearest molecules, as given in Table \ref{tLC} for both the simple cubic (SC) and hexagonal close
packed (HCP) or face-centred cubic (FCC) lattices.

\begin{table}[ht] 
  \caption{Lattice constants.} 
  \label{tLC} 
  \centering 
  \begin{tabular}{l l l} 
    \hline\hline 
    Lattice  & $c_\name{r} $ & $c_\name{a} $   \rule{0pt}{2.6ex}\\ 
    \hline 
    SC      & $24.8085962$ & $33.6076959$     \rule{0pt}{2.6ex}\\ 
    HCP/FCC & $48.5275208$ & $57.8156842$    \\ 
    \hline 
  \end{tabular} 
\end{table}

The virial equation, obtained by setting ${\dd^2I / \dd t^2 = 0}$ in the Lagrange-Jacobi identity, contains terms
proportional to $M$ except the gravity term, which is proportional to $M^{5/3}$. Dividing the equation by $M$ we can
separate $M^{2/3}$, yielding,
\begin{equation}
  M^{2/3} = \left(375 \over 4 \pi \rho\right)^{1/3} {1 \over G m} 
  \left[k_\name{B}T + 4 c_\name{r}\epsilon\left({\sigma^3\rho\over m}\right)^4 
    -2 c_\name{a} \epsilon\left({\sigma^3\rho\over m}\right)^2
  \right],
\end{equation}
which is indeed positive if the sum of the terms inside the brackets are also positive.  However the term proportional
to $c_\name{a}$ is negative, and when large introduces the possibility that no positive solution for $M^{2/3}$ exists.
Solving the bracket terms equal to zero for $\left(\sigma^3\rho/m\right)^2$, we have then a quadratic equation for this
term, which gives the solutions for the densities $\rho_0$ for which $M$ vanishes,
\begin{equation}
  \left({\sigma^3\rho_0\over m}\right)^2 = {1 \pm \sqrt{1-4\,{c_\name{r} \over c_\name{a}^2}{k_\name{B}T \over \epsilon}} 
   \over 4 \,{c_\name{r}\over c_\name{a} } }.
\end{equation}
The $+$ and $-$ solutions are real non-negative when the term inside the square root above is non-negative.  The maximum
temperature at which $M$ can vanish is thus,
\begin{equation}
  \label{eTmax} 
  {k_\name{B} T_\name{max} \over \epsilon }= {c_\name{a}^2 \over 4 c_\name{r} } . 
\end{equation}
Since the critical temperature for a phase transition in the absence of gravity is about $\epsilon/k_\name{B}$, we see
that $T_\name{max}$ can be substantially larger than the critical temperature.  With the values given in Table 1 for the
constants $c_\name{a}$ and $c_\name{r}$, the maximum temperature below which gravity combined with molecular forces
enhances fragmentation are $T_\name{max}=414.3\,\name{K}$ and $626.8,\name{K}$ for the SC and HCP/FCC lattices,
respectively, which are $11.38$ and $17.22$ larger than $\epsilon/k_\name{B}$.

The corresponding density $\rho_\mathrm{f}$ at which fragmentation can occur at \textit{arbitrarily small mass} is
\begin{equation} 
  \label{erho0}
  \rho_\mathrm{f} = {m \over \sigma^3} \sqrt{ c_\name{a} \over 4 c_\name{r} }\ , 
\end{equation}
which is of the order of the individual molecule density $m / \sigma^3$. 

The Lagrange-Jacobi identity therefore allows us to predict a density $\rho_\mathrm{f}$ and a maximum temperature
$T_\name{max}$ below which a homogeneous sphere made of LJ molecules can find a gravitational equilibrium at an
arbitrarily small mass: in a way, this provides a simple model, an explanation, for the reason of forming substellar ice
clumps out of a cold self-gravitating gas.  For H$_2$ molecules arranged as SC or HCP/FCC lattice, we find
$\rho_\mathrm{f}= 73.8\,\name{kg\,m^{-3}}$ and $69.2\,\name{kg\,m^{-3}}$, respectively.  These densities are of the
order or slightly below the solid or liquid H$_2$ density ($\approx 80\,\name{kg\,m^{-3}}$).

At temperatures $T > T_\name{max}$ the isothermal equilibrium curves have a minimum mass when $dM/d\rho|_\name{T}=0$,
which is expressed as
\begin{equation} 
  M_\name{min}^2 = K {\epsilon^3 \sigma^3\over G^3 m^4}
  {c_\name{a}^{11/2}\over c_\name{r}^{5/2}} { \left[ {11 \over 5 }\tau
      - \left( 1 + \sqrt{ 1+{11 \over 25 }\tau }\right) \right]^{3} \over \sqrt{1+\sqrt{1
        +{11\over 25}\tau }}}
\end{equation}
where $\tau = T/T_\name{max}$ and
\begin{equation} 
  K = {81\over 2\pi} \left(5 \over 11\right)^{11/2} \approx 0.1686476934 \ .
\end{equation}

Figure \ref{fMT} shows the minimum mass as a function of the temperature for HCP/FCC H$_2$ homogeneous spheres.  It is
rising very steeply at $T \gtrsim 627\,\name{K}$, but rising much slower after $\sim 1000\,\name{K}$.  Interestingly the
mass range includes all the masses below terrestrial planet masses for temperatures below H$_2$ dissociation.

\subsubsection{Condensed bodies}

Different condensed and uncondensed bodies can be identified using the Lagrange-Jacobi identity, as summarized in
Table~\ref{tLJ}.

\begin{table}[ht] 
  \caption{Condensed and uncondensed bodies in a LJ fluid using the Lagrange-Jacobi identity.} 
  \label{tLJ} 
  \centering 
  \begin{tabular}{l | l} 
    \hline\hline 
    Dominating Terms & Name   \rule{0pt}{2.6ex}\\ 
    \hline 
    $2E_\name{kin} + 6 E_\name{pot, LJ, a}$& ``molecular gas''     \rule{0pt}{2.6ex}\\ 
    $12 E_\name{pot, LJ, r} + 6 E_\name{pot, LJ, a}$& ``comet''\\ 
    $12 E_\name{pot, LJ, r} + E_\name{pot,G}$& ``rocky planetoid''\\ 
    $2E_\name{kin} + E_\name{pot,G}$& ``gaseous planetoid''\\ 
    \hline 
  \end{tabular} 
\end{table}

Figure \ref{fvirial-ratios} shows some $M(\rho,T)$ equilibrium curves of gravitating homogeneous sphere whose molecules
are arranged on a HCP/FCC lattice, which is able to represent the most compact sphere lattice at high density.  At low
density, on the left of the diagram, the equilibrium is principally fixed by the attractive part of molecules and their
temperature, and the exact lattice structure is irrelevant. This is the domain of uncondensed ``molecular gas''.

On the right, there is an accumulation curve at approximately $\rho \approx 100 \, \name{kg\,m^{-3}}$ representing
purely condensed H$_2$ bodies mainly balancing gravity with the repulsive part of the molecular interaction. The area
between the dashed line, connecting the minimum of the isotherms with $T > T_\name{max}$, and the accumulation curve is
the domain of ``rocky planetoids''.

At temperatures below $T_\name{max}=626.8\,\name{K}$, the equilibrium curves plunge to arbitrarily small masses along two
regimes: the left vertical asymptotes represent the limit gaseous bodies, and the right asymptotes the condensed (solid
or liquid) bodies called ``comets''. The area of the ``comets'' lies between the dotted line, connecting the elbows of
the isotherms with $T < T_\name{max}$, and the dashed line.

At high temperature and low density, to the left of both the dotted and dashed line and above the isotherm lines, lie
the ``gaseous planetoids''. These bodies' equilibrium is fixed principally by gravity and temperature.

Of course real astrophysical bodies are not homogeneous, in practice in the small mass regime one can expect bodies made
of a mixture of condensed state in the core surrounded by a less dense gaseous atmosphere, which could be calculated by
solving the hydrostatic equilibrium, see \citet{pfenniger_cold_2004}, where some models of isothermal bodies made of
H$_2$ containing a solid core and a gaseous atmosphere have been discussed.

\subsection{Gravitational instability} 
\label{ssGI} 

The linearized wave equation for the isentropic density perturbation $\rho_\name{per}$ of a self-gravitating fluid of
density $\rho$ is the following \citep{jeans_stability_1902,weinberg_gravitation_1972}:
\begin{equation} 
  \frac{\partial^2\rho_\name{per}}{\partial t^2} - \left(\frac{\partial P}{\partial \rho}\right)_\name{s} \nabla^2\rho_\name{per} =
  4\pi G  \rho\rho_\name{per}\ .
\end{equation} 
Its solution is a superposition of modes of the form $\exp\left(i(\vec k\cdot \vec x - \omega t) \right)$, where
$\omega$ is the frequency and $k = {2\pi}/{\lambda}$ the wavenumber, and where $\lambda$ is the wavelength. This leads
to the instability condition \citep{jeans_stability_1902},
\begin{equation} 
  \label{eo} 
  \omega^2 =  \left(\frac{\partial P}{\partial \rho}\right)_\name{s} k^2 - 4\pi G\rho < 0 \ .
\end{equation}
When fulfilled the modes are real-exponential, therefore unstable.  The classical Jeans criterion reads,
\begin{equation} 
  \lambda > 
  \lambda_\name{J} \equiv  
  \sqrt{ \left( \frac{\partial P}{\partial \rho} \right)_\name{s} \frac{\pi}{G\rho}
  } \ .
\label{eJeans} 
\end{equation}
Therefore the effective gravitational instability is directly dependent on the EOS of the medium.

\subsubsection{Ideal gas}

In most textbooks about Jeans instability only the ideal gas is discussed.  In the case of an ideal monoatomic gas, the
pressure derivative term is
\begin{equation}
  \left(\frac{\partial P}{\partial \rho}\right)_\name{s} = \gamma\frac{k_\name{B}T}{m} \ ,
\label{eig} 
\end{equation}
with the adiabatic index $\gamma = {5}/{3}$. This value is always positive, which leads to the classical, ideal
monoatomic gas Jeans criterion for instability,
\begin{equation} 
  \lambda > \lambda_\name{J} = \sqrt{\frac{\pi \gamma k_\name{B}T}{G\rho m}} \ .
  \label{eJeansideal} 
\end{equation}

\subsubsection{Fluid with phase transition}

\begin{figure}[t]
  \resizebox{\hsize}{!}{\includegraphics{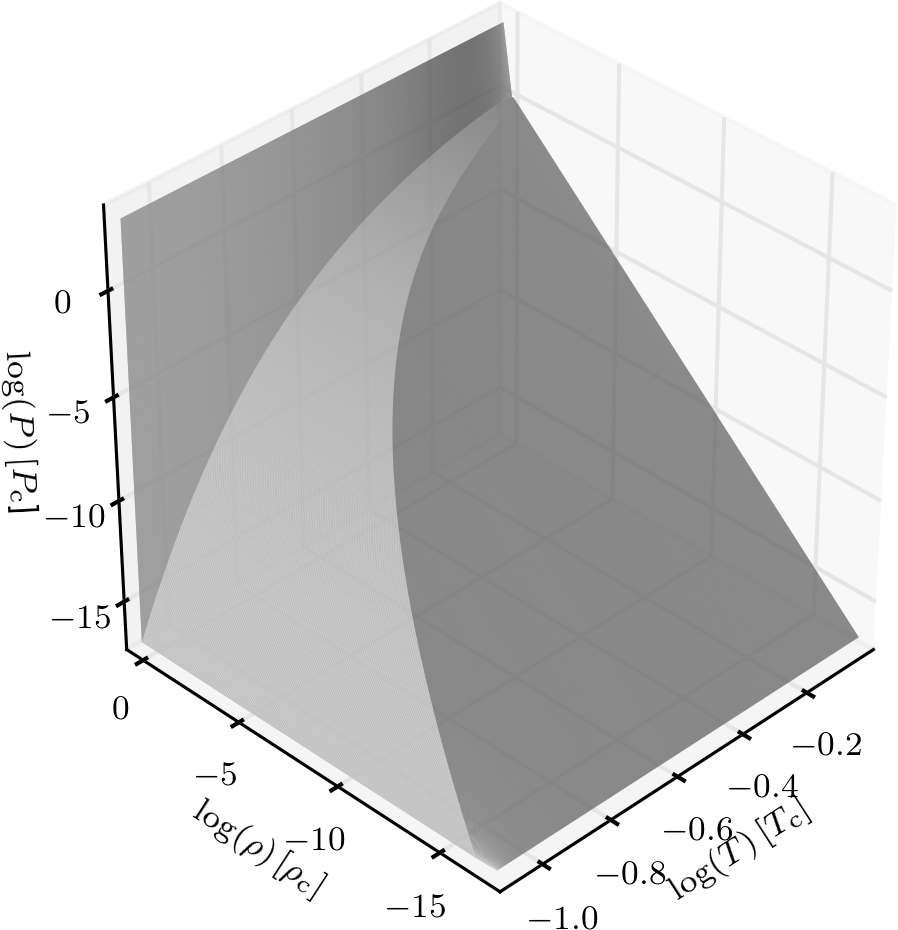}}
  \caption{van der Waals EOS, including the Maxwell construct. }
  \label{fgravin}
\end{figure}

When considering a fluid presenting a phase transition the term $\left(\partial P /\partial \rho \right)_\name{s}$ may
differ from that of an ideal gas.  As seen before, in the case of the van der Waals EOS, the Maxwell construct must be
used, which may considerably change the combined stability of the fluid (Sect.  \ref{ssVdW}).  In the presence of a
phase transition, the EOS gradient is strongly modified, in particular $(\partial P / \partial \rho)_\name{T} = 0$.
Using
\begin{equation}
  \left(\partial P \over \partial \rho\right)_\name{s} = {c_\name{P}\over c_\name{v}}\left(\partial P \over \partial \rho\right)_\name{T}
\end{equation}
with $c_\name{P}$ and $c_\name{v}$ both finite values, we find $\left(\partial P /\partial \rho \right)_\name{s} = 0$.
Therefore, a self-gravitating fluid in a phase transition is also automatically gravitationally unstable.

Figure \ref{fgravin} shows the 3D representation of the EOS including the Maxwell construct. Clearly the curved
triangular region, the phase transition region (on the left) has a very different gradient than the almost ideal-gas
region at low density and high temperature, or the condensed phase region at high density.  It is remarkable that a
temperature drop by one order of magnitude from the critical temperature leads to a drop in the critical pressure by 14
orders of magnitude. For example for H$_2$ at 3\,K the critical pressure is $6.4\cdot 10^{-9}\,\unit{Pa}$.
    
\section{Molecular dynamics} \label{sMD}

For all simulations, the Large-Scale Atomic/Molecular Massively Parallel Simulator (LAMMPS) is used
\citep{plimpton_fast_1995}. The LAMMPS software is a state of the art and widely used single and multi-processor code in
chemistry, material sciences, and related fields. Its abilities to quickly compute short and long-range forces and the
possibility to use a multi-timescale integrator make it a suitable tool to perform our simulations.

\subsection{Potential Solver} 
\label{ssPS} 

The LAMMPS code has a wide range of force fields. For the simulations in this article we use only the LJ and
self-gravitational potentials.

Since the straight calculation cost for exact pairwise interactions is $\name{O}(N^2)$, approximate but still accurate
methods are implemented that make the calculations possible at a much lower cost.  A cut-off radius $r_\name{c}$ is used
for the short-range forces. As the attractive part of the LJ potential drops with $r^{-6}$, it is calculated only for
neighbour particles within $r_\name{c}$.  The neighbours for each particle are found using a Verlet neighbour list. This
list is created with a radius of $r_\name{n} = r_\name{c} + r_\name{s}$, $r_\name{s}$ being an extra ``skin'' distance
to avoid the recalculation of the Verlet list at every time-step. This cut-off method is $\name{O}(N)$ and therefore
linearly scales with the number of particles.

The gravitational potential drops with $r^{-1}$ so the long range interactions cannot be ignored. It is calculated using
the Particle$^3$-Mesh (P$^3$M) method \citep{hockney_computer_1981}. The gravitational potential is split in short-range
and long-range parts in Fourier space \citep{ewald_berechnung_1921}:
\begin{equation} 
  \hat{\Phi} = \hat{\Phi}_\name{SR}\left(1 - \name{exp}(-k^2r_\name{s}^2)\right) + 
  \hat{\Phi}_\name{LR}\,\name{exp}(-k^2r_\name{s}^2) \ , 
\end{equation} 
where $r_\name{s}$ is the splitting distance. The potential $\Phi_\name{SR}$ is calculated at the same time as the LJ
potential, using the same Verlet list. The potential $\Phi_\name{LR}$ is calculated in Fourier space using a fast
Fourier transform (FFT).

As LAMMPS is designed for the simulation of chemical substances not far from terrestrial conditions, and self-gravity is
too weak to be of any influence, no specific self-gravity module is provided. For that reason, a tweak is used by using
the Coulomb's potential module for an electric field,
\begin{equation} 
  \Phi_\name{C} = \frac{Cq_iq_j}{\epsilon r} \ ,
\end{equation} 
where $C$ is the interaction constant, $\epsilon$ the dielectric constant, and $q_{i, j}$ the charges of the molecules.
As $C$ is fixed in LAMMPS, to correctly calculate the gravitational potential we set the constants as follows:
\begin{eqnarray} 
  \epsilon &=& -1 \ ,
  \\ q_i &=& \sqrt{\frac{G}{C}}m_i \ .
\end{eqnarray}

\subsection{Time integration} 
\label{ssTI} 

The time integration is done using the symplectic leapfrog scheme. The drift and kick operators, 
\begin{eqnarray} 
  D(\Delta t) &\equiv& x(t + \Delta t) = x(t) + \Delta t \dot{x}(t)    \ ,\\ 
  K(\Delta t) &\equiv& \dot{x}(t + \Delta t) = \dot{x}(t) + \Delta t \ddot{x}(t) \ , 
\end{eqnarray} 
are applied according to the sequence
$K\left(\frac{\Delta t}{2}\right)\, D(\Delta t) \, K\left(\frac{\Delta t}{2}\right)$ at each elementary time-step. For
short-range force, the time-step is constant throughout the simulation and set as $10^{-2}$ -- $10^{-3}$, which is the
typical interaction timescale during nearest-neighbour molecular interactions.  This is reasonable since the repulsive
interaction and the finite kinetic energy prevent strongly varying accelerations between particles.

Small changes of individual particle positions can significantly change the short-range forces, which is why a very
small time-step is required.  But these small changes have almost no influence on long-range forces. For this reason,
there is no need to calculate the long-range forces at every time-step.

In regard to short-range calculations, long-range calculations are an order of magnitude more costly per step but always
need the same amount of calculation time: creation of the density-map with a fifth order interpolation procedure, a FFT
solution of the potential, and the same interpolation procedure for finding the accelerations.

With the ``reversible reference system propagator algorithm'' (rRESPA), a multiple timescale integrator is available for
LAMMPS that enables us to reduce the number of long-range force calculation \citep{tuckerman_reversible_1992}. It
enables time integration in up to four hierarchical levels, but only two are needed for the simulations we present.  The
rRESPA time-integration-scheme looks as follows:
\begin{equation} 
  K_\name{LR}\left(\frac{\Delta t}{2}\right) \,
  \left[K_\name{SR} \left(\frac{\Delta t}{2n_\name{SR}}\right)\,  D\left(\frac{\Delta t}{n_\name{SR}}\right)\,
    K_\name{SR} \left(\frac{\Delta t}{2n_\name{SR}}\right)
  \right]^{n_\name{SR}}
  K_\name{LR}\left(\frac{\Delta t}{2}\right) \ ,
\end{equation} 
where $n_\name{SR}$ is the number of short-range iterations and set as $10$ -- $100$, $K_\name{LR}$ is the kick operator
using the long-range (FFT) accelerations, and $K_\name{SR}$ the kick operator using the short-range accelerations.

\subsection{Super-molecules} 
\label{ssSM} 

The maximum number of particles that can be simulated on today's supercomputers is of the order of $10^9$ -- $10^{10}$
particles. Even using such a huge number of molecules would only result in a fluid with a total mass of a few
femtograms. It is obvious that gravity has no effect on this kind of fluid. For that reason, the concept of
super-molecules is introduced. This concept is well established in galactic dynamics simulation, where
super-stars weigh typically $10^4 - 10^6\,\unit{M}_\odot$, and in cosmology where super-WIMPS may weigh as
much as $\sim 10^{67}$ GeV$ c^{-2}$ particles. Two-body relaxation or diffusion due to the low number of super-particles
is negligible, provided the simulation time is not too large, depending on the specific problem.  Practice has shown
that this kind of an approximation is valid in galactic dynamics for instance if the simulation length is of order 100
dynamical times \citep{binney_galactic_2008}.

The basic principle of the concept is that each super-molecule consists of $\eta$ molecules, and  its mass is therefore,
\begin{equation} 
  m_{\mathrm{SM}} = \eta\, m_\mathrm{M} \ .
 \label{emSM} 
\end{equation} 
To have the same properties of a super-molecule fluid as for a normal fluid, every term of the virial Equ. (\ref{eVir}) has
to be transformed such that the ratios of the terms are invariant. The kinetic energy term reads
\begin{equation} 
  2 E_\name{kin} = N_\mathrm{M} m_\name{M}\langle{v_\name{M}^2}\rangle = N_\mathrm{SM} m_\name{SM}\langle{v_\name{SM}^2}\rangle \ .
\end{equation}
Since the total mass is constant, $N_\mathrm{M} m_\name{M} = N_\mathrm{SM} m_\name{SM}$, the velocity dispersion of
molecules and super-molecules is also the same,
\begin{equation} 
  \langle{v_\mathrm{SM}^2}\rangle = \langle{v_\name{M}^2}\rangle \ .
\end{equation}
From the two LJ terms, Equ.\ (\ref{eELJr}--\ref{eELJa}) remain the same if 
\begin{eqnarray}
  {\epsilon_\name{M}\sigma_\name{M}^{12} \over m_\name{M}^5} &=& {\epsilon_\mathrm{SM}\sigma_\mathrm{SM}^{12} \over m_\mathrm{SM}^5}\ ,\\
  {\epsilon_\name{M}\sigma_\name{M}^6 \over m_\name{M}^3} &=& {\epsilon_\mathrm{SM}\sigma_\mathrm{SM}^6 \over m_\mathrm{SM}^3} \ .
\end{eqnarray}
Solving these two equations and using Equ.\ (\ref{emSM}), we derive 
\begin{eqnarray}
  \epsilon_\mathrm{SM} &=& \eta \,\epsilon_\name{M} \ ,\\ 
  \sigma_\mathrm{SM} &=& \sqrt[3]{\eta}\sigma_\name{M} \ . 
\end{eqnarray}
The gravitational energy Equ.\ (\ref{eEgrav}) does not change since it does not depend on super-molecule properties.

It is important to ensure that the gravitational force between two super-molecules remains small compared to the
corresponding LJ force within the short-range force cut-off radius. This is assured by setting
$\vec{F}_\name{G}(r_\name{c}) \ll \vec{F}_\name{LJ}(r_\name{c})$ with $r_c$ being the cut-off radius. This leads to the
following constraint:
\begin{equation} 
  \label{exglj}
  \eta^\frac{2}{3} \ll {24\epsilon_\name{M}\sigma_\name{M} \over G\,m_\name{M}^2} 
  \left( r_\name{c,SM}^{-5} -  2r_\name{c,SM}^{-11} \right) \ ,
\end{equation}
where $r_\name{c,SM} = (r_\name{c} /\sigma_\name{SM})$. Using a cut-off radius of $r_\name{c} = 4\,\sigma$ and hydrogen
super-molecules, we find a maximum super-molecule mass equal to $5.7 \cdot 10^{-6}\,M_\oplus$, meaning at least
$1.7\cdot 10^5$ particles are required for simulating an Earth-mass, virialized, H$_2$-condensed body.

\subsection{Ice clump detection} 
\label{ssIG} 

\begin{figure}[t] 
  \resizebox{\hsize}{!}{\includegraphics{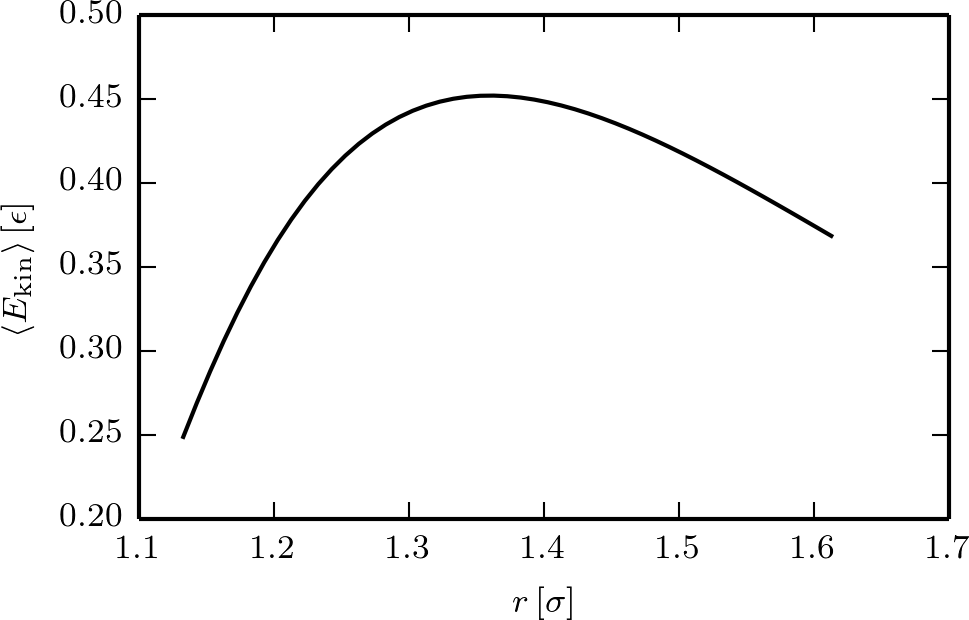}}
  \caption{Mean kinetic energy as a function of the maximum displacement in an ice clump.}
  \label{fdMax}
\end{figure}

The LJ potential has its minimum at $r_m = 2^{1/6}\sigma$. If an ice clump is at absolute zero temperature, all bound
molecules would have this distance from at least one other molecule. But, as the molecules in a clump are vibrating, the
maximum binding distance between two molecules is generally larger than $r_m$.

The simplest clump consists of two molecules. Their mean kinetic energy as a function of the maximum displacement can be
calculated numerically (Fig.~\ref{fdMax}).

The mean kinetic energy rises up to the maximum value of $r_{\sigma, \name{max}} = 1.3625$, after which it drops
again. It can be assumed that all stable clumps have a binding distance below this value. The distance constraint for
two molecules to be bound is therefore:
\begin{equation} 
  r_\sigma \leq r_{\sigma, \name{max}} \ .
\end{equation}
Using $r_{\sigma, \name{max}}$ as the threshold distance, LAMMPS provides a list of clumps at fixed time intervals.

\section{Simulations} 
\label{sS} 

\begin{table}[ht] 
  \caption{Parameters of the super-molecule test simulations.} 
  \label{tSM} 
  \centering 
  \begin{tabular}{l l l l} 
    \hline\hline 
    Name & $n/n_\name{cr}$ & $T/T_\name{cr}$ & $N_\name{SM}$\rule{0pt}{2.6ex}\rule[-1.2ex]{0pt}{0pt}\\ 
    \hline 
    SM15 & $0.006$ & $1.5$ & $4^3$ -- $200^3$\rule{0pt}{2.6ex}\\ 
    SM30 & $0.006$ & $3.0$ & $4^3$ -- $200^3$\\ 
    SM60 & $0.006$ & $6.0$ & $4^3$ -- $200^3$\\ 
    \hline 
    SM04 & $0.1$ & $0.4$ & $25^3$ -- $200^3$\rule{0pt}{2.6ex}\\
    SM06 & $0.1$ & $0.4$ & $25^3$ -- $200^3$\\
    \hline
    SF\tablefootmark{a} & $0.1$ & $0.1$ & $50^3$ -- $160^3$\rule{0pt}{2.6ex}\\
    \hline
  \end{tabular} 
  \tablefoot{All simulations are without gravity and use the basic KDK time integration scheme.\\
    \tablefoottext{a}{External gravity $a=0.1\,(L / \tau^2)$}} 

  \medskip

  \caption{Parameters of the one-phase fluid simulations.} 
  \label{tOP} 
  \centering 
  \begin{tabular}{l l l l l } 
    \hline\hline 
    Name & $n/n_\name{cr}$ & $T/T_\name{cr}$ & $N_\name{SM}$ & $\gamma_\name{J}$\rule{0pt}{2.6ex}\rule[-1.2ex]{0pt}{0pt}\\ 
    \hline 
    OP0    & $10^{-2}$ & $1.25$ & $100^3$ & $0$ \rule{0pt}{2.6ex}\\ 
    OPGw   & $10^{-2}$ & $1.25$ & $100^3$ & $0.5$\\ 
    OPGs   & $10^{-2}$ & $1.25$ & $50^3$ -- $160^3$& $1.25$\\ 
    \hline 
  \end{tabular} 
  \tablefoot{ OP0 without gravity and using the KDK time integration scheme. All gravitational simulations use the P$^3$M 
    gravitational solver and the rRESPA time scheme.} 

  \medskip

  \caption{Parameters of the phase transition simulations.} 
  \label{tTP} 
  \centering 
  \begin{tabular}{l l l l l} 
    \hline\hline 
    Name & $n/n_\name{cr}$ & $T/T_\name{cr}$ & $N_\name{SM}$ & $\gamma_\name{J}$\rule{0pt}{2.6ex}\rule[-1.2ex]{0pt}{0pt}\\
    \hline 
    \emph{OPGs}	&$10^{-2}$ & $1.25$ & $50^3$ -- $160^3$ & $0$, $0.5$, $1.25$\rule{0pt}{2.6ex}\\ 
    \hline 
    PT1-1	& $10^{-1}$ & $0.1$ & $50^3$, $100^3$ & $0$, $0.5$, $G=G_\name{OPGs}$\rule{0pt}{2.6ex}\\ 
    PT2-1    & $10^{-1}$ & $0.2$ &$50^3$, $100^3$ & $0$, $0.5$, $G=G_\name{OPGs}$\\ 
    PT3-1   & $10^{-1}$ & $0.3$ & $50^3$, $100^3$ & $0$, $0.5$, $G=G_\name{OPGs}$\\ 
    \hline 
    PT1-2	& $10^{-2}$ & $0.1$ & $50^3$, $100^3$ & $0$, $0.5$, $G=G_\name{OPGs}$\rule{0pt}{2.6ex}\\ 
    PT2-2    & $10^{-2}$ & $0.2$ &$50^3$ -- $100^3$ & $0$, $0.5$, $G=G_\name{OPGs}$\\ 
    PT3-2   & $10^{-2}$ & $0.3$ & $50^3$ -- $100^3$ & $0$, $0.5$, $G=G_\name{OPGs}$\\ 
    PT5-2	& $10^{-2}$ & $0.5$ & $50^3$, $100^3$ & $0$, $0.5$, $G=G_\name{OPGs}$\\ 
    PT7-2    & $10^{-2}$ & $0.7$ &$50^3$, $100^3$ & $0$, $0.5$, $G=G_\name{OPGs}$\\
    PT9-2   & $10^{-2}$ & $0.9$ & $50^3$, $100^3$ & $0$, $0.5$, $G=G_\name{OPGs}$\\ 
    \hline 
    PT1-3   & $10^{-3}$ & $0.1$ & $50^3$, $100^3$ & $0$, $0.5$, $G=G_\name{OPGs}$\rule{0pt}{2.6ex}\\ 
    PT2-3   & $10^{-3}$ & $0.2$ & $50^3$, $100^3$ & $0$, $0.5$, $G=G_\name{OPGs}$\\ 
    PT3-3   & $10^{-3}$ & $0.3$ & $50^3$, $100^3$ & $0$, $0.5$, $G=G_\name{OPGs}$\\ 
    \hline 
  \end{tabular}
  \tablefoot{All simulations in two different runs: without gravity and perturbation, using basic KDK time integration scheme, 
    and with gravity and perturbation, using the P$^3$M gravitational solver and the rRESPA time scheme.} 
\end{table}

\subsection{Units}
A fluid is defined by the number of super-molecules $N_\name{SM}$, the initial velocity distribution (temperature), the
number density, and the strength of the gravitational potential. All simulation properties are molecule-independent, the
initial temperature and density are measured as a factor of the critical values $T_\name{cr}$ and $n_\name{cr}$, and the
gravitational potential is measured as the factor $\gamma_\name{J} = G / G_\name{J}$ of the ideal-gas Jeans gravity
$G_\name{J}$, defined as
\begin{equation}
  G_\name{J} = {\pi\gamma k_\name{B} T_0\over n m^2 L^2}
\end{equation} 
with the box side $L = (N_\name{SM}/n)^{1/3}$. It is interesting to note that $G_\name{J}$ is proportional to
temperature and inversely proportional to density. For all simulations, the time unit is defined as the average time of
a particle crossing the box
\begin{equation} 
  \tau = \frac{L}{V} \ ,
\end{equation}
with $V^2 = \sum_i {v_x}_i^2/N$.

The distance at which the gravitational and LJ forces are equal is denoted as
\begin{equation}
  \label{exGLJ}
  x_\name{GLJ} \equiv F_\name{G}(x_\name{GLJ}) = F_\name{LJ}(x_\name{GLJ}) \ .
\end{equation}

\subsection{Super-molecule concept tests} 
\label{ssSMS}
 
\subsubsection{Potential energy}

\begin{figure}[t] 
  \resizebox{\hsize}{!}{\includegraphics{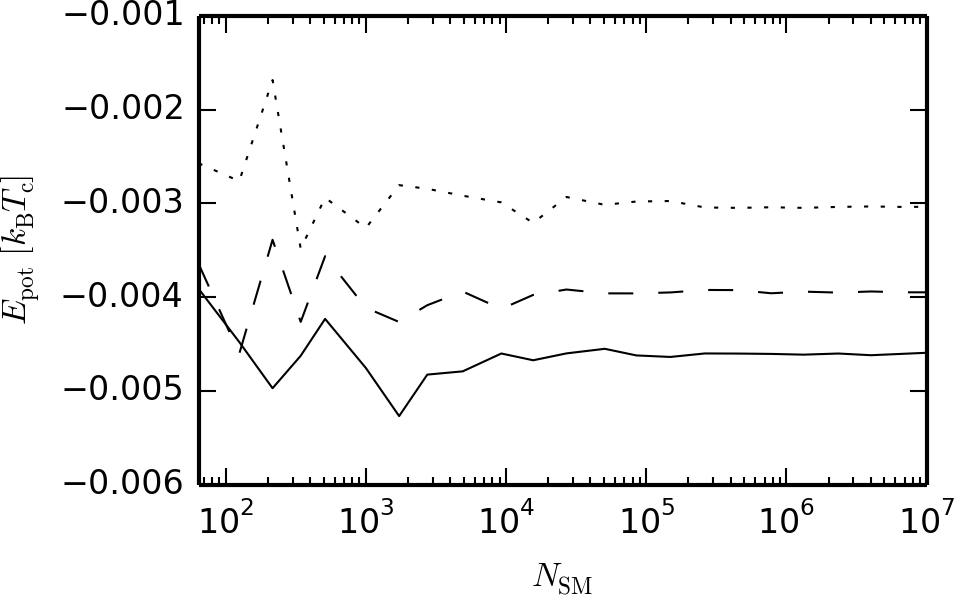}}
    \caption[LJ potential energy as a function of the number of super-molecules at $t = 1\tau$. Solid line: SM15; dashed
    line: SM30; dotted line: SM60.]{LJ potential energy as a function of the number of super-molecules at $t = 1\tau$.
    	Solid line: SM15$^a$; dashed line: SM30$^a$; dotted line:
    	SM60$^a$. \\ $^{(a)}$ See Table \ref{tSM}} 
  \label{fEp}
\end{figure}

A fluid in a cubic box with periodic boundary conditions and initial constant density of $n = 0.006\,n_\name{cr}$ is
simulated with LAMMPS.  Three different initial Maxwellian velocity distributions with temperature $T=1.5$, $3.0$ and
$6.0\,T_\name{cr}$, and a number of particles from $N_\name{SM}=30^3$ -- $200^3$ are used.  Table \ref{tSM} shows the
parameters of the different simulations.

Figure \ref{fEp} shows the LJ potential energy of those three fluids with respect to the number of super-molecules. One
can see fluctuations in the simulations with low $N_\name{SM}$, but the result becomes reasonable when
$N_\name{SM} > 10^4$ and a satisfactory convergence is obtained if $N_\name{SM} > 10^5$.  Therefore all the subsequent
simulations are performed with $N_\name{SM} > 10^5$.

\subsubsection{Cluster percentage}

\begin{figure}[t] 
  \resizebox{\hsize}{!}{\includegraphics{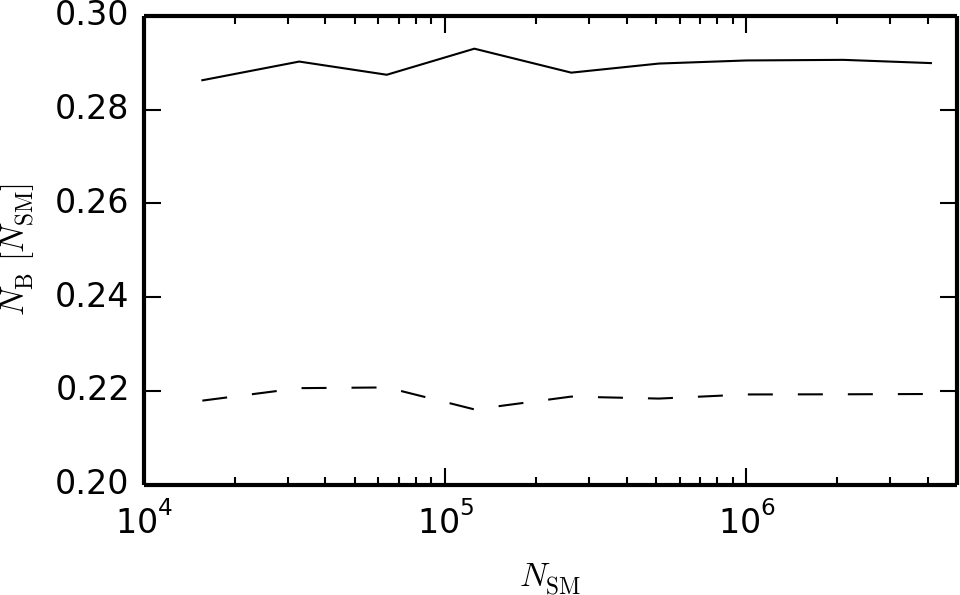}}
  \caption[Fraction of bound molecules as a function of the number of super-molecules at $t = 5\tau$. Solid line: SM04,
  dashed line: SM06.]{Fraction of bound molecules as a function of the number of super-molecules at $t = 5\tau$. Solid
  	line: SM04$^a$, dashed line: SM06$^a$. \\ $^{(a)}$ See Table \ref{tSM}}
  \label{fSMnb}
\end{figure}

High density, low temperature fluids are simulated during $5 \tau$ at various number of super-molecules. These fluids
will form ``comets'' and allow us to compare the final percentage of bound molecules. Table \ref{tSM} shows the
parameters of the different simulations.

Figure \ref{fSMnb} shows the fraction of bound molecules with respect to the number of super-molecules. All simulations
reach the same percentage, independent of the number of super-molecules.

\subsubsection{Precipitation in an external  field}

\begin{figure}[t] 
  \resizebox{\hsize}{!}{\includegraphics{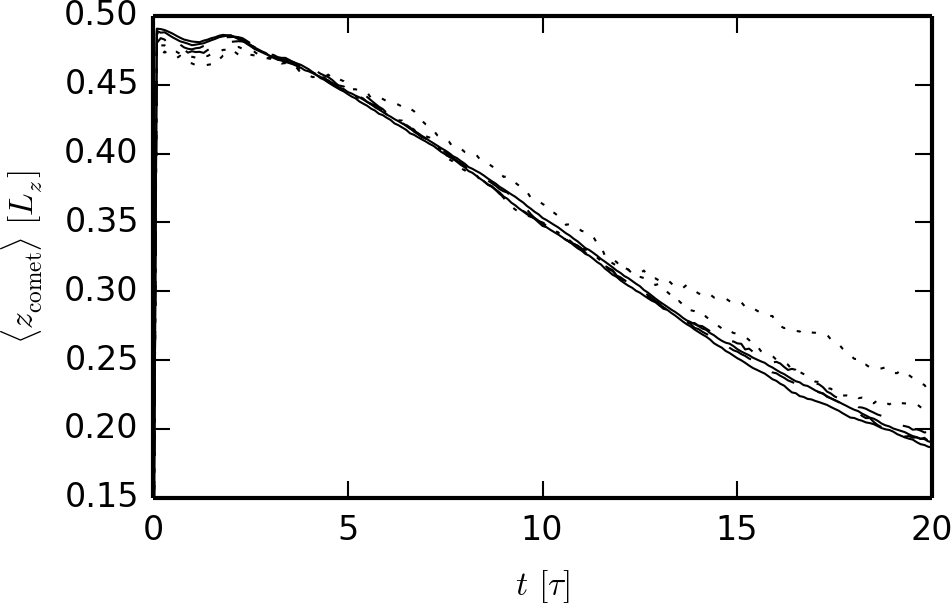}} 
  \caption[$z$-Coordinate of the centre of mass of ``comets'' as a function of time of the SF simulation. $N_\name{SM} =
  50^3$ and $64^3$ (dotted line), $80^3$ and $100^3$ (dashed line), $128^3$ and $160^3$ (solid line).]{$z$-Coordinate of
  	the centre of mass of ``comets'' as a function of time of the SF$^a$ simulation. $N_\name{SM} = 50^3$ and $64^3$
  	(dotted line), $80^3$ and $100^3$ (dashed line), $128^3$ and $160^3$ (solid line). \\ $^{(a)}$ See Table \ref{tSM}}
  \label{fsnowZ}
\end{figure}

An external acceleration is applied in the $-z$ direction to a fluid within a box with reflecting boundary
conditions. The fluid is very cold and rather dense (see Table \ref{tSM}) and is therefore forming ice grains, or
``comets''. Because of acceleration and Archimedes's principle, the ``comets'' fall to the bottom and stay there,
similar to snowfall on Earth.

The purpose of this simulation is to test the timescaling of the precipitation as function of the number of particles.
Figure \ref{fsnowZ} shows the $z-$coordinate of the centre of mass of ``comets'' as a function of time. The curves are
very similar, with a slight over-estimation of the collapse time for the smallest simulation.  Therefore the
precipitation phenomenon timescale is not strongly dependent on the mass resolution.

\subsection{Fluid with perturbation} 
\label{ssfwp} 

To illustrate the behaviour of a homogeneous fluid perturbed by a plane sinusoidal wave, a velocity perturbation in the
$x$ direction is introduced with wavelength $\lambda$ equal to the box side $L$.  Starting with a Maxwellian
distribution $\vec{v}_i$ of velocities for the homogeneous unperturbed case, the $x$-component of the velocities is
perturbed in such a way as to conserve energy. The perturbed $x$-velocity component for each particle $i$ reads,
\begin{equation} 
  {v^\prime_x}_i = \alpha \left[{v_x}_i + \beta V\sin(\omega x_i)\right] \ ,
  \label{evs} 
\end{equation} 
with $\omega =2\pi/L$.  The correction factor $\alpha\leq 1$ is used to keep the same kinetic energy in the $x$
direction, with
\begin{equation} 
  \alpha^2= \frac{\sum_i{v^2_x}_i}{\sum_i\left[{v_x}_i+ \beta V\sin(\omega x_i)\right]^2} \ .
\end{equation} 
$\beta$ determines the strength of the perturbation, but is supposed to be small enough to be in the linear regime of
perturbations. In the present simulations, $\beta = 0.01$.

\subsection{One-phase fluid} 
\label{ssOPS} 

To study the reaction of a pure one-phase ideal-gas fluid far from the phase transition but with the above plane-wave
perturbation, the initial temperature is taken above the critical value ($T_0 = 1.25\,T_\name{cr}$) and the initial
density well below the critical value ($n_0 = 10^{-2}\,n_\name{cr}$). Three different cases are studied: without
gravity, weakly self-gravitating below the classical ideal-gas Jeans criterion with $\gamma_\name{J} = 0.5$, and
sufficiently self-gravitating above the ideal-gas Jeans criterion with $\gamma_\name{J} = 1.25$. The simulations are
performed with $N_\name{SM}$ ranging from $50^3$ to $160^3$.  Table \ref{tOP} shows the parameters of the different
simulations.

\subsubsection{Time evolution}

\begin{figure}[t] 
  \resizebox{\hsize}{!}{\includegraphics{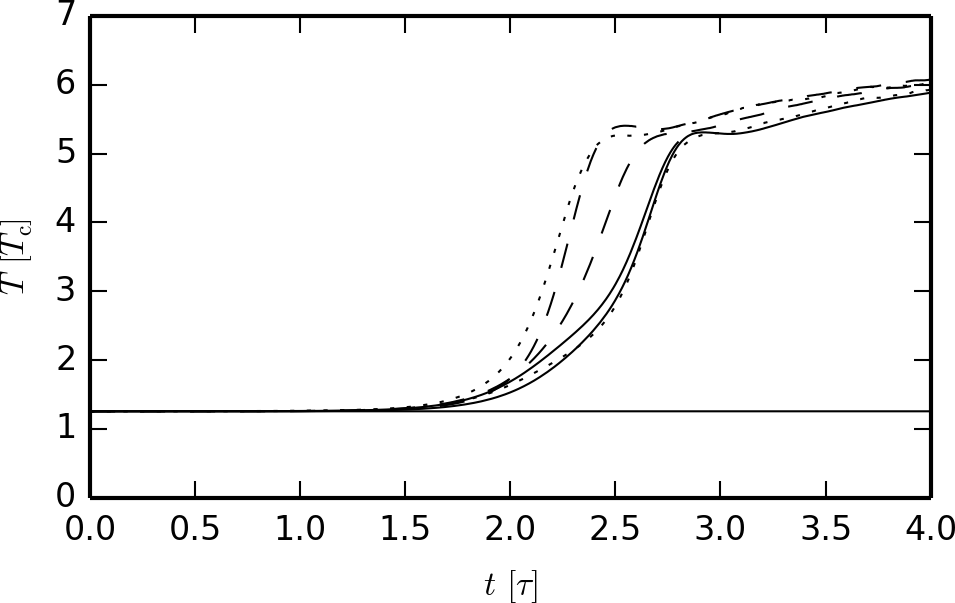}}
  \caption[Temperature of the one-phase fluid simulations OP0 (straight solid line) and OPGs as a function of time.
  $N_\name{SM} = 50^3$ and $64^3$ (dotted line), $80^3$ and $100^3$ (dashed line), $128^3$ and $160^3$ (solid line).]
  {Temperature of the one-phase fluid simulations OP0$^a$ (straight solid line) and OPGs$^a$ 
  	as a function of time. $N_\name{SM} = 50^3$ and $64^3$ (dotted line), $80^3$ and $100^3$ (dashed line), $128^3$ and
  	$160^3$ (solid line). \\ $^{(a)}$ See Table \ref{tOP}}
  \label{fG}
\end{figure}

\begin{figure}[t] 
  \resizebox{\hsize}{!}{\includegraphics{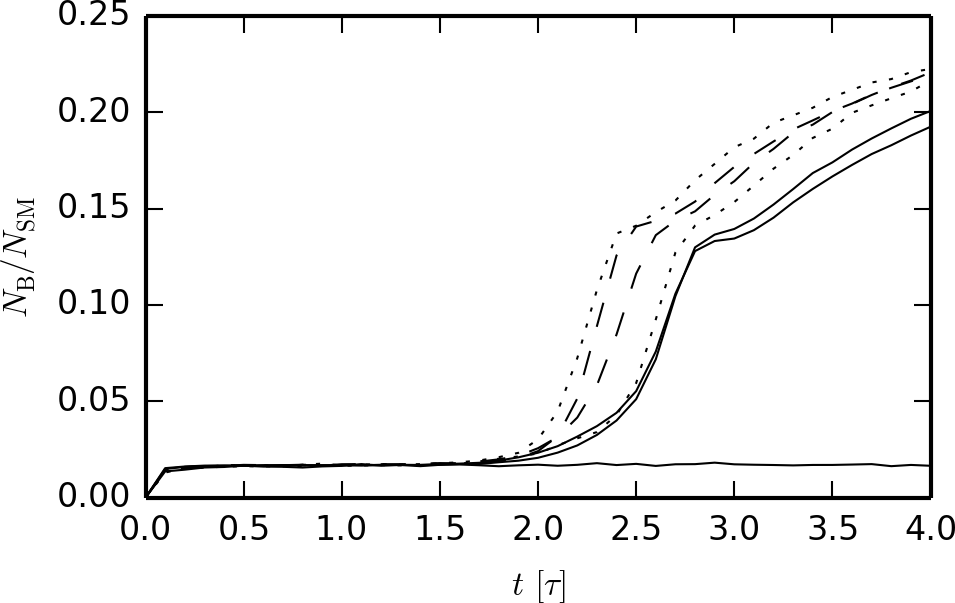}}
  \caption{Fraction of bound molecules in ``comets'' of the one-phase fluids OP0 (straight solid line) and OPGs as a
    function of time. $N_\name{SM} = 50^3$ and $64^3$ (dotted line), $80^3$ and $100^3$ (dashed line), $128^3$ and
    $160^3$ (solid line).}
  \label{fGc}
\end{figure}

\begin{figure}[t] 
  \resizebox{\hsize}{!}{\includegraphics{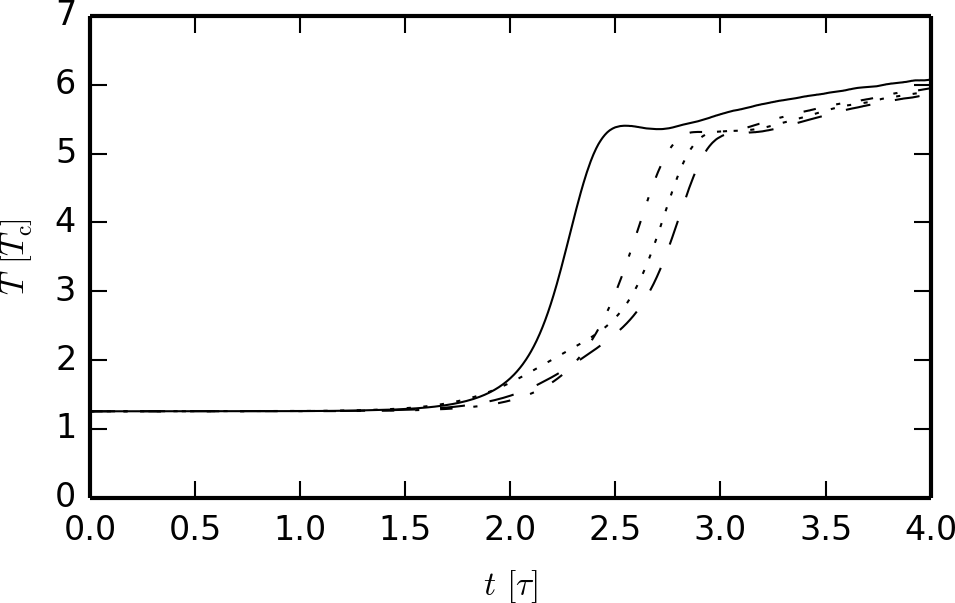}}
  \caption{Temperature of the one-phase fluid simulations OPGs as a function of time. $N_\name{SM} = 100^3$ with four
    different random seeds.}
  \label{fG100}
\end{figure}

\begin{figure}[t] 
  \resizebox{\hsize}{!}{\includegraphics{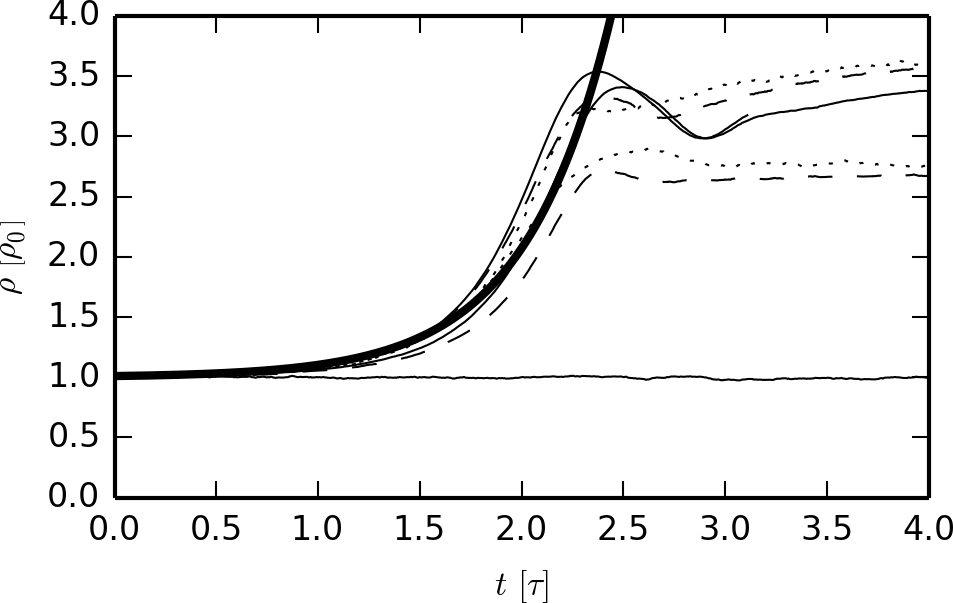}}
  \caption{Density at $x=(0.5\pm0.05)L$ of the one-phase fluids OP0 (straight solid line) and OPGs as a function of
    time.  Bold line: analytic solution for ideal gas.  $N_\name{SM} = 50^3$ and $64^3$ (dotted line), $80^3$ and
    $100^3$ (dashed line), $128^3$ and $160^3$ (solid line).}
  \label{fGd}
\end{figure}

\begin{figure}[t] 
  \resizebox{\hsize}{!}{\includegraphics{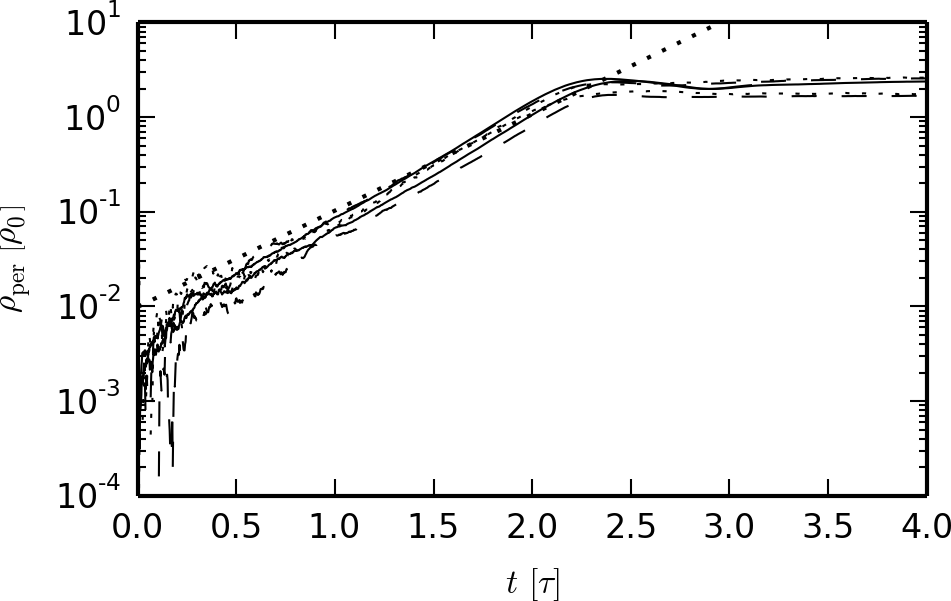}}
  \caption{Evolution of perturbation density. Straight dotted line: analytic solution for ideal gas.
    $N_\name{SM} = 50^3$ and $64^3$ (dotted line), $80^3$ and $100^3$ (dashed line), $128^3$ and $160^3$ (solid line).}
  \label{fexp}
\end{figure}

As the simulations conserve total energy, the formation of ``comets'', implying a decrease of potential energy, can be
measured by the mean temperature (i.e., kinetic energy) of the system. This is equivalent to the release of latent heat
when a gaseous fluid condenses.

Figures \ref{fG} and \ref{fGc} display the temperature and the ``comet'' percentage evolutions of the one-phase fluid
simulations: OP0 without gravity and $N_\name{SM} = 100^3$ super-molecules and the sufficiently self-gravitating fluid
OPGs with $N_\name{SM}$ ranging from $50^3$ to $160^3$. The weakly self-gravitating fluid of simulation OPGw is not
shown since it is identical to the no gravity case OP0.

For the fluid without gravity OP0 and the weakly self-gravitating fluid OPGw, no particular effect is observed. The
percentage of ``comets'' rises to $\lesssim 2\%$, which can be attributed to the initial fluctuations in the
distribution. The small perturbation introduced in the $x$-direction does not change the nature of the fluid; its
temperature and ``comet'' percentage remains the same.

The sufficiently self-gravitating OPGs fluids, on the other hand, do change. Their temperatures and cluster percentages
are rising steeply showing a substantial latent heat release. All simulations reach the same asymptotic temperature
$\sim 6 T_\name{cr}$, but different initial conditions (i.e., random seeds) yield slightly different behaviours during
the collapse. This can also be observed in Fig.~\ref{fG100} where four identical parameter simulations but different
random seeds are run with $N_\name{SM} = 100^3$, leading to a time difference of $\sim 0.5\tau$ for reaching the
asymptotic upper value.

Figure \ref{fGd} shows the density increase around the centre of the perturbation in the range $x = (0.5\pm0.05)L$ and
compares them to the analytic solution for an ideal gas. The simulations follow the ideal-gas solution with a slight
dispersion up to $t \approx 2.5\tau$, where a density rise is not possible anymore because of the repulsive LJ force;
the density declines for a short while because of the collapse rebound down to a minimum at $t \approx 2.7\tau$. This
density local minimum corresponds with the break in the temperature increase visible in Fig.~\ref{fG}. Subsequently, the
temperature growth is much slower.

Figure \ref{fexp} shows the evolution of density in the simulations, and the predicted growth rate line for an ideal gas
subject to Jeans' instability (Sect. \ref{ssGI}). All curve slopes correspond initially to the predicted Jeans' growth
rate. The initial steep growth and the small time delay are caused by the initial noise in the particle
distribution. Thus, a certain time is needed for the perturbation to overcome the noise.

\subsubsection{``Comet'' mass distribution}
\label{ssCMD}

\begin{figure}[t] 
  \resizebox{\hsize}{!}{\includegraphics{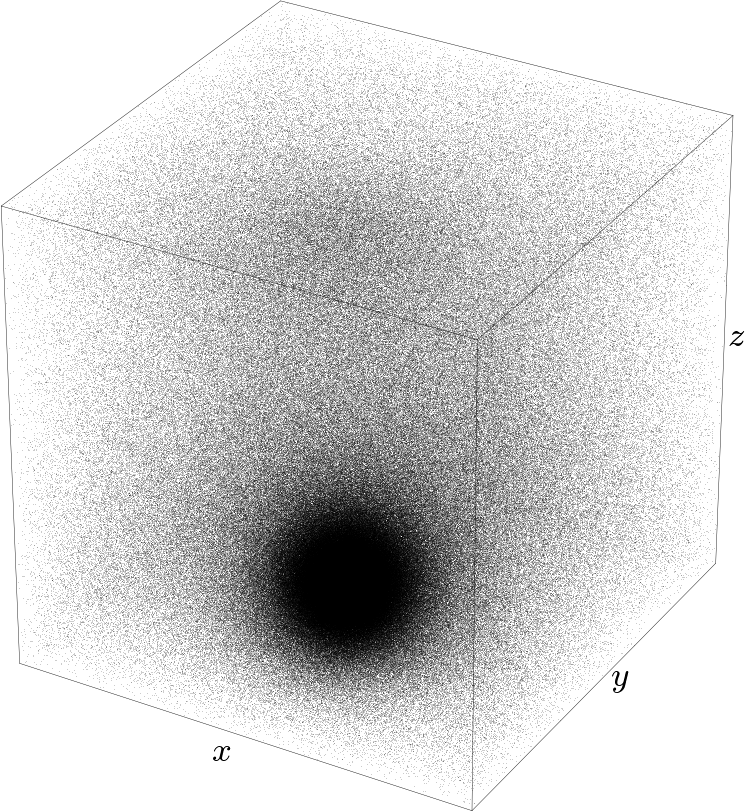}}
  \caption{3D-view of simulation OPGs with $N_\name{SM} = 100^3$at $t = 4\tau$.}
  \label{fG122s}
\end{figure}

\begin{figure}[t] 
  \resizebox{\hsize}{!}{\includegraphics{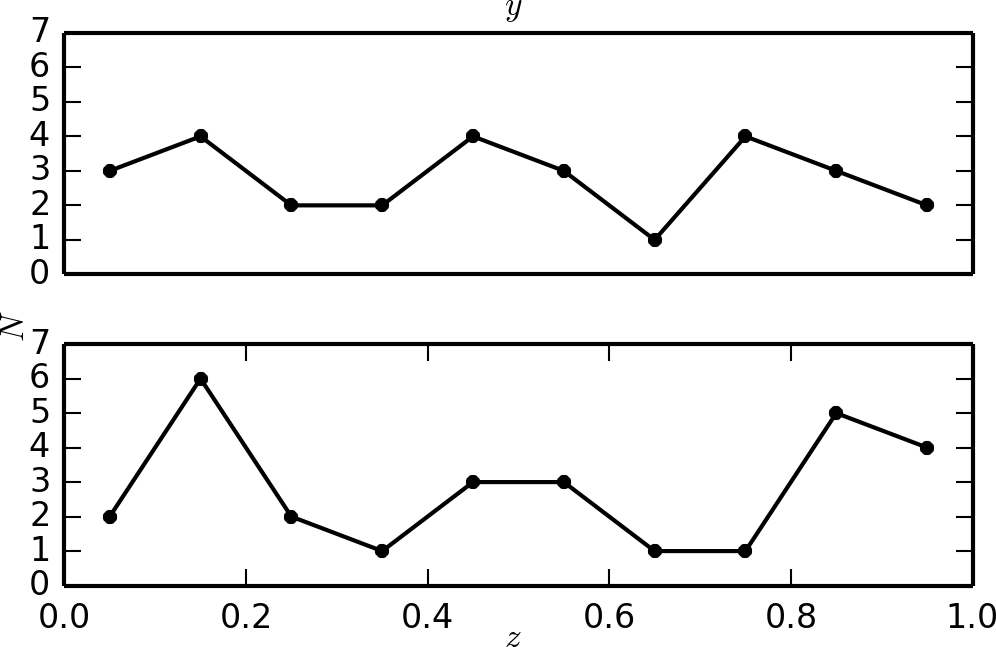}}
  \caption{Distribution of y- and z-coordinates of ``planetoid'' centre of mass on a 10x10 grid; all OPGs simulations at
    $t = 2.5\tau$.}
  \label{fPDist}
\end{figure}

\begin{figure}[t] 
  \resizebox{\hsize}{!}{\includegraphics{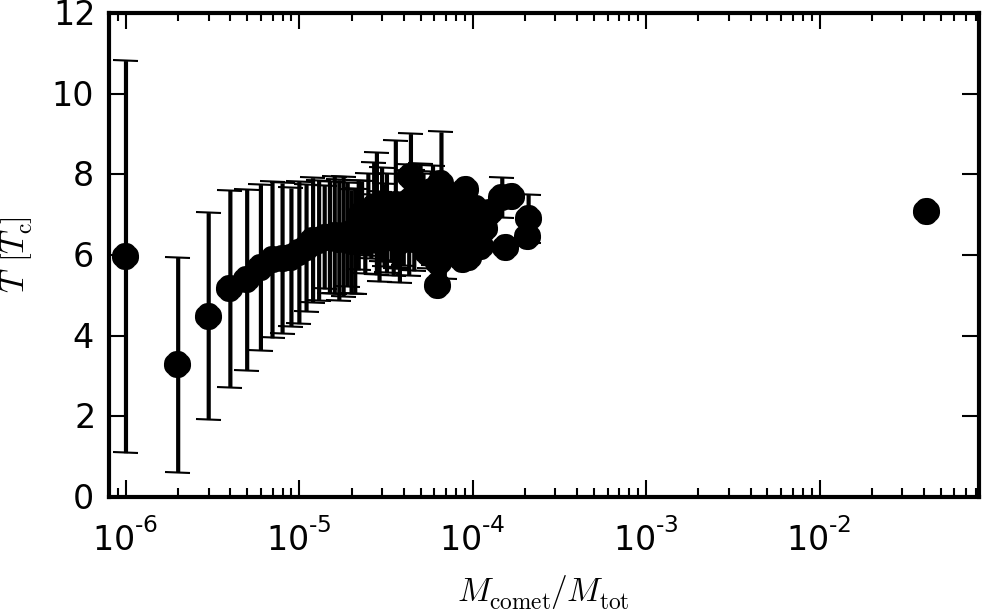}}
  \caption{Temperature distribution of unbound molecules and ``comets'' as a function of ``comet'' mass of OPGs with
    $N_\name{SM} = 100^3$ at $t = 4\tau$.}
  \label{fOPGT}
\end{figure}

\begin{figure*}[p]
  \centering
  \includegraphics[width=18cm]{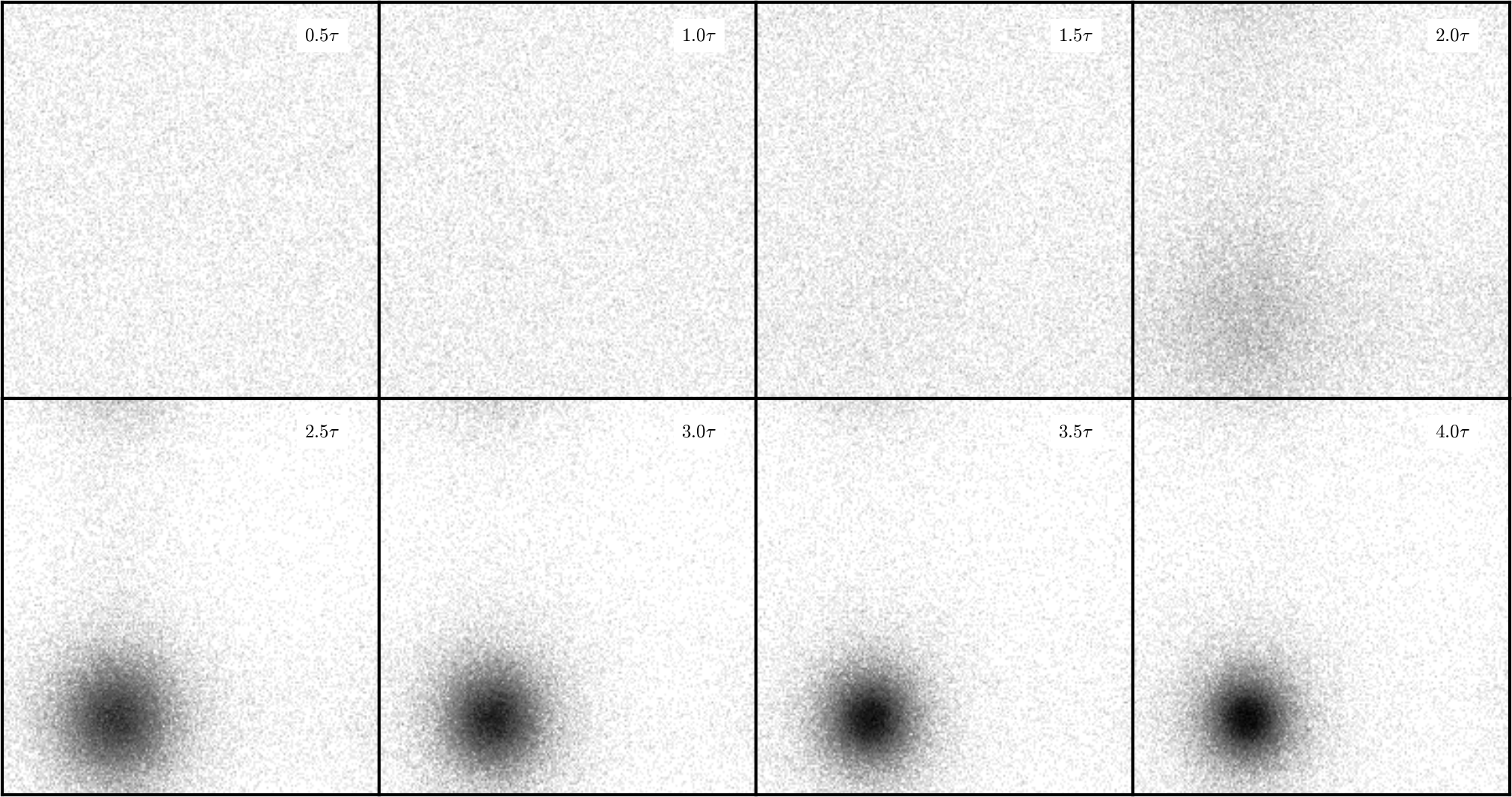}
  \caption{Snapshot time sequence of the simulation OPGs with $N_\name{SM} = 100^3$, showing the smooth formation of a
    ``planetoid''.  The slice size is $0.025 \times 1\times 1 \,L^3$ at $x=(0.5\pm0.0125)L$.}
  \label{fGs}
 
  \medskip

  \centering
  \includegraphics[width=18cm]{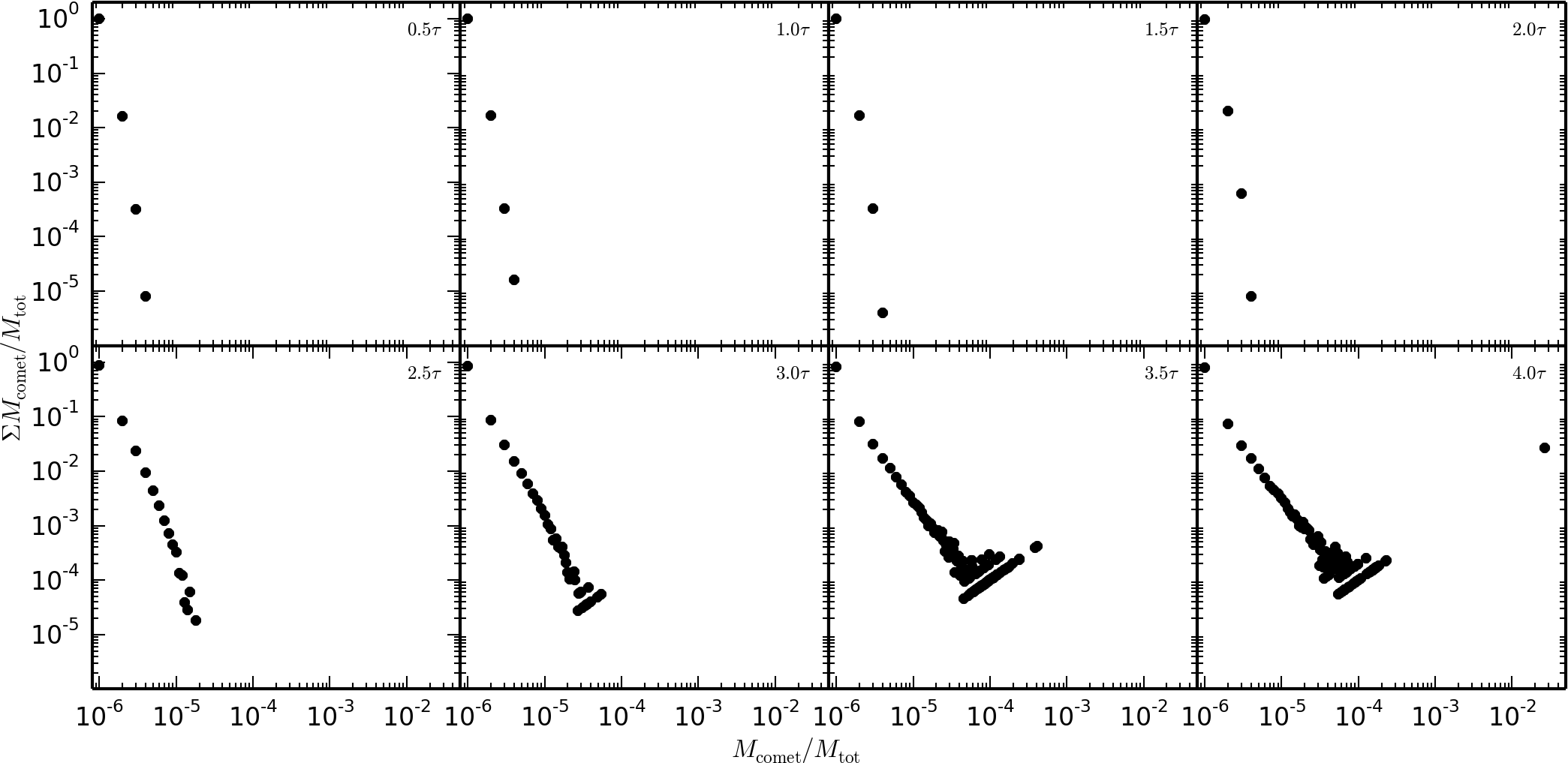}
  \caption{Snapshot ``comet'' mass-distribution sequence of the simulation OPGs with $N_\name{SM} = 100^3$.}
  \label{fGh}
\end{figure*}

Figure \ref{fGs} shows snapshots and Fig.~\ref{fGh} the ``comet'' mass distribution at different times. During the
temperature rise, the ``comets'' are distributed according to a power law,
\begin{equation}
  \label{ePC}
  {\dd\log\left(\sum M_\name{comet}\right) \over \dd\log\left(M_\name{comet}\right)} = 
  \xi_\name{c}\,\,\name{for}\, M_\name{comet} \ll M_\name{tot} \ ,
\end{equation}
and no ``planetoid'' is formed. The power-law index is very steep at the beginning with $\xi_\name{c} \approx -10$, but
quickly decreases as it reaches a value of $\xi_\name{c} \approx -2.5$ at the end.

After the temperature increase break at $t\approx 2.7\tau$, the power-law distribution remains for small ``comets'', but
one bigger ``gaseous planetoid'' with over $1\%$ of the total mass forms. It is shown in Fig.~\ref{fG122s}. The
spherical body seen at $t \leq 3\tau$ in the snapshots is uncondensed gas, which is why it does not figure in the
``comet'' mass distribution (see Sect. \ref{ssIG} for a discussion how condensed matter is identified).

A second power law:
\begin{equation}
  \label{ePP}
  {\dd\log\left(\sum M_\name{comet}\right) \over \dd\log\left(M_\name{comet}\right)} = 
  \xi_\name{p}\,\,\name{for}\, M_\name{comet} \gg M_\name{SM}
\end{equation}
on the right side of the diagram describes the ``comets'' and ``planetoids''. Each body occurs typically only once with
our limited $N_\name{SM}$, and $\xi_\name{p} = 1$. Equation (\ref{ePC}) describes the mass distribution of small bodies
where gravity is negligible whereas Equ. (\ref{ePP}) describes large bodies where gravity is weak, but cannot be
neglected.

The first high-density plane is created thanks to the plane perturbation parallel to the $yz$-plane at $x=0.5$. One can
observe in Fig.~\ref{fGs} at $t=2\tau$ that two filaments form, along the $y$- and $z$- axes, connecting the
``planetoid'' with itself thanks to the periodic boundary conditions. They are aligned with the mesh, but not primarily
due to grid effects, but because these lines are the shortest distance between the ``planetoid'' and its periodic
images. As can be seen in Fig. \ref{fPT1-3s}, diagonal filaments are also possible using the second shortest distance.

The position of the ``planetoid'' is, thanks to the plane-wave perturbation, situated in the $x = 0.5L$ plane, the $y$
and $z$ positions are random and depend on the initial condition and vary with every simulation as shown in
Fig. \ref{fPDist}.

Figure \ref{fOPGT} shows the temperature distribution in the ``comets'' as function of mass. As the number of large
``comets'' with the same number of super-molecules is small or even one, no error bars can be seen for the larger
``comets'' and the ``gaseous planetoid''.

The broad Maxwellian velocity distribution is seen for the unbound molecules ($M_\name{comet} = 10^{-6}$). The small
``comets'' (2 or more particles) clearly have a lower temperature and dispersion than the average temperature of the
system. This is due to the fact that in order for two super-molecules to cluster together, a third one is needed, and
this takes away some of their kinetic energy. The larger a ``comet'' is, the closer its temperature is to the system
average temperature.  On longer formation time the ``planetoids'' temperature tends towards the system average
temperature, as expected.

\subsubsection{Scaling}

\begin{figure}[t] 
  \resizebox{\hsize}{!}{\includegraphics{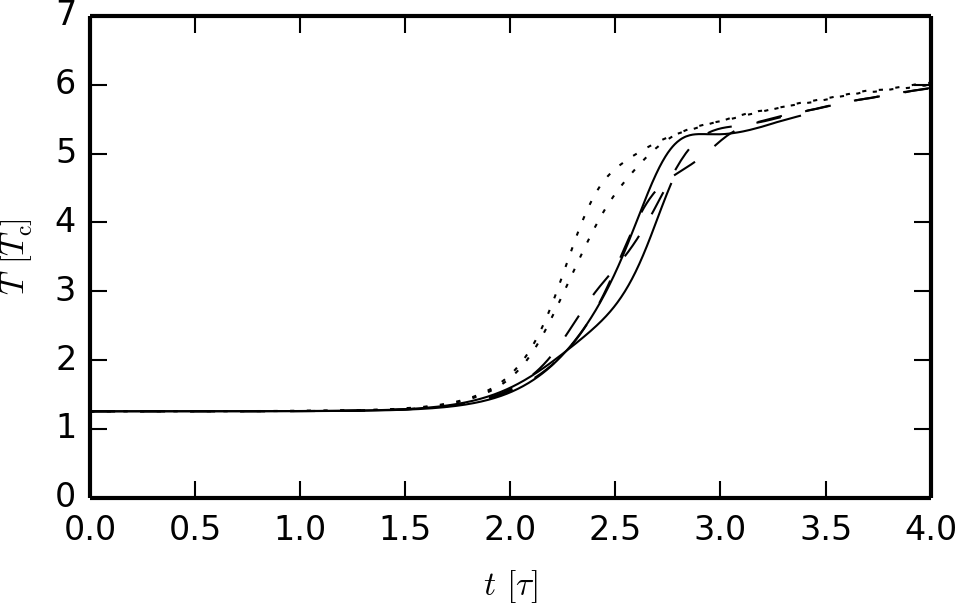}}
  \caption{Mean temperature over four one-phase fluid simulations OPGs with different random seeds as a function of
    time.  $N_\name{SM} = 50^3$ and $64^3$ (dotted line), $80^3$ and $100^3$ (dashed line), $128^3$ and $160^3$ (solid
    line).}
  \label{fGmean}
\end{figure}

\begin{figure}[t] 
  \resizebox{\hsize}{!}{\includegraphics{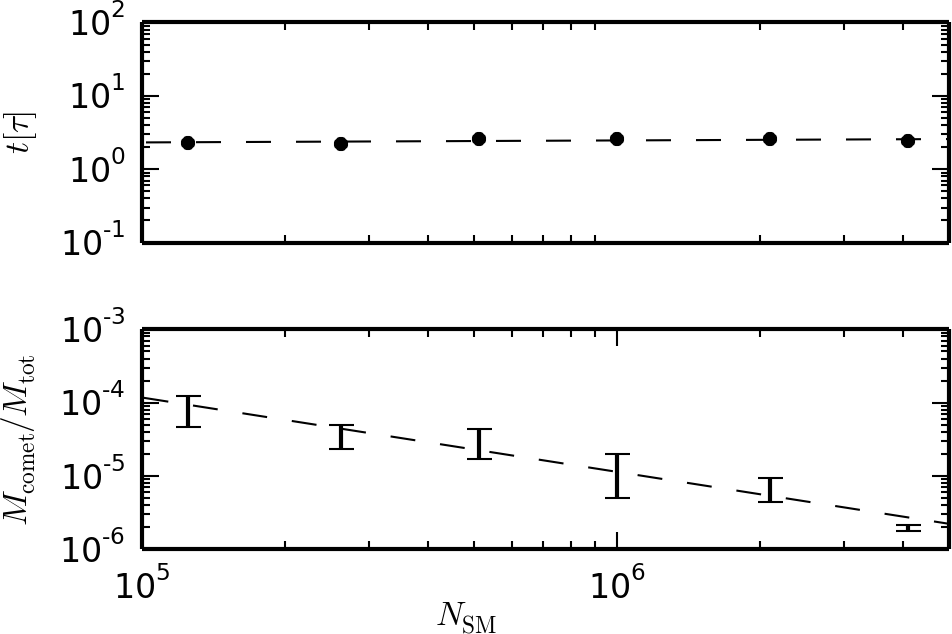}}
  \caption{Time and size of the largest ``comet'' at the appearance of the turning point.}
  \label{fturn}
\end{figure}

Figure \ref{fGmean} shows the mean temperature value over four simulations with different random seeds. One can see that
the two smallest simulations with $N_\name{SM} = 50^3$ and $64^3$ are starting to collapse earlier than the other
simulations, which are very similar. This can be explained by the fact that these simulations have very low
$x_\name{GLJ}$ values (Equ. \ref{exGLJ}), $3.65$ and $4.03\,r_\sigma$. The first value is in fact lower than the cut-off
radius of $4\,r_\sigma$ and fails to satisfy Equ.~(\ref{exglj}), the other one just slightly above it. In these
simulations, the gravitational forces are strong even in short-range intermolecular interactions, the simulations are
therefore overemphasizing the gravitational effect.

\begin{figure*}[p]
  \centering
  \includegraphics[width=18cm]{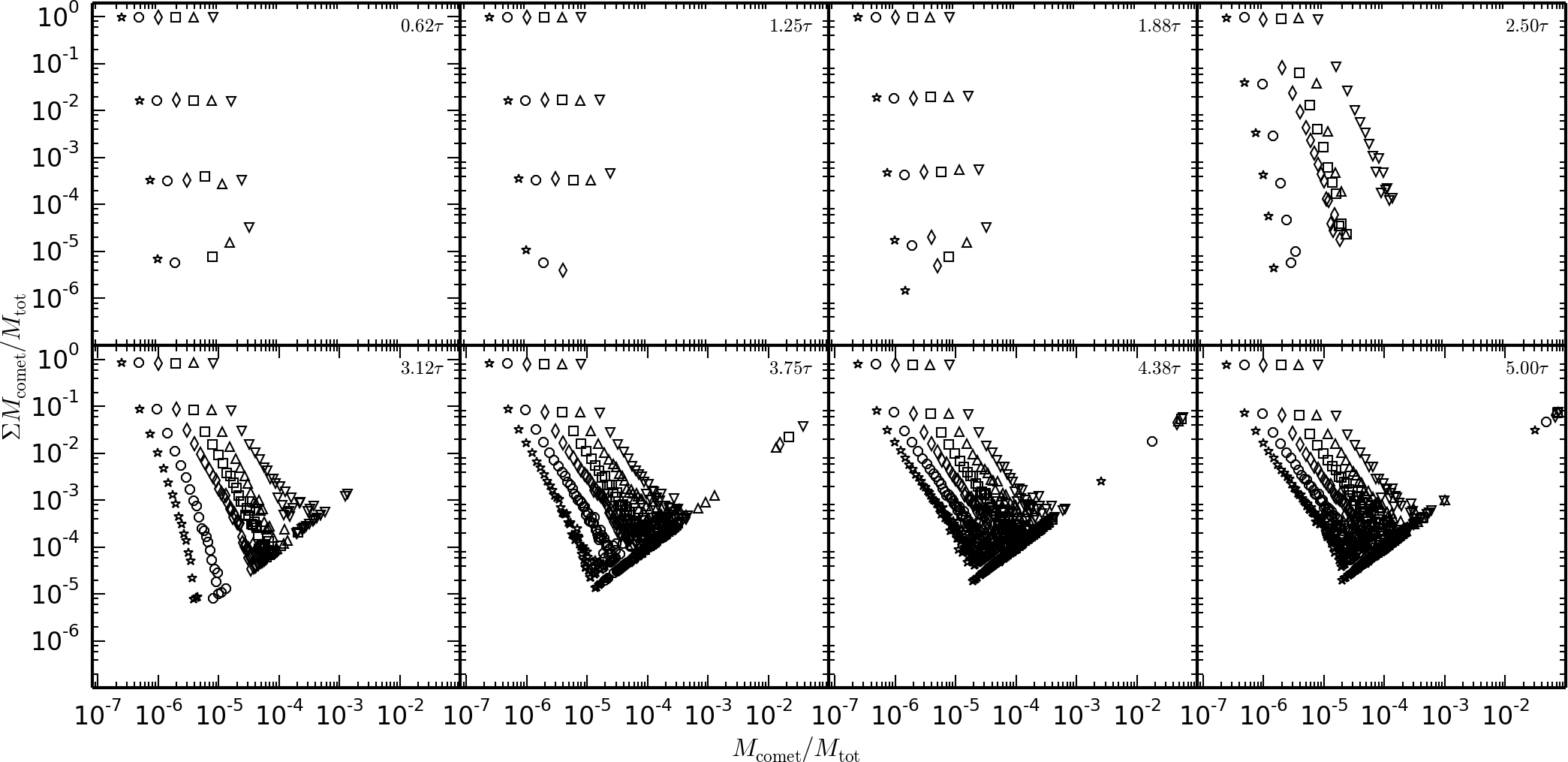}
  \caption{Snapshot ``comet'' mass-distribution sequence of all OPGs simulations with $N_\name{SM}$ equal to the
    triangle pointing down: $50^3$, triangle pointing up: $64^3$, square: $80^3$, diamond: $100^3$, circle: $128^3$,
    star: $160^3$.}
  \label{fGall}
  \medskip \centering
  \includegraphics[width=18cm]{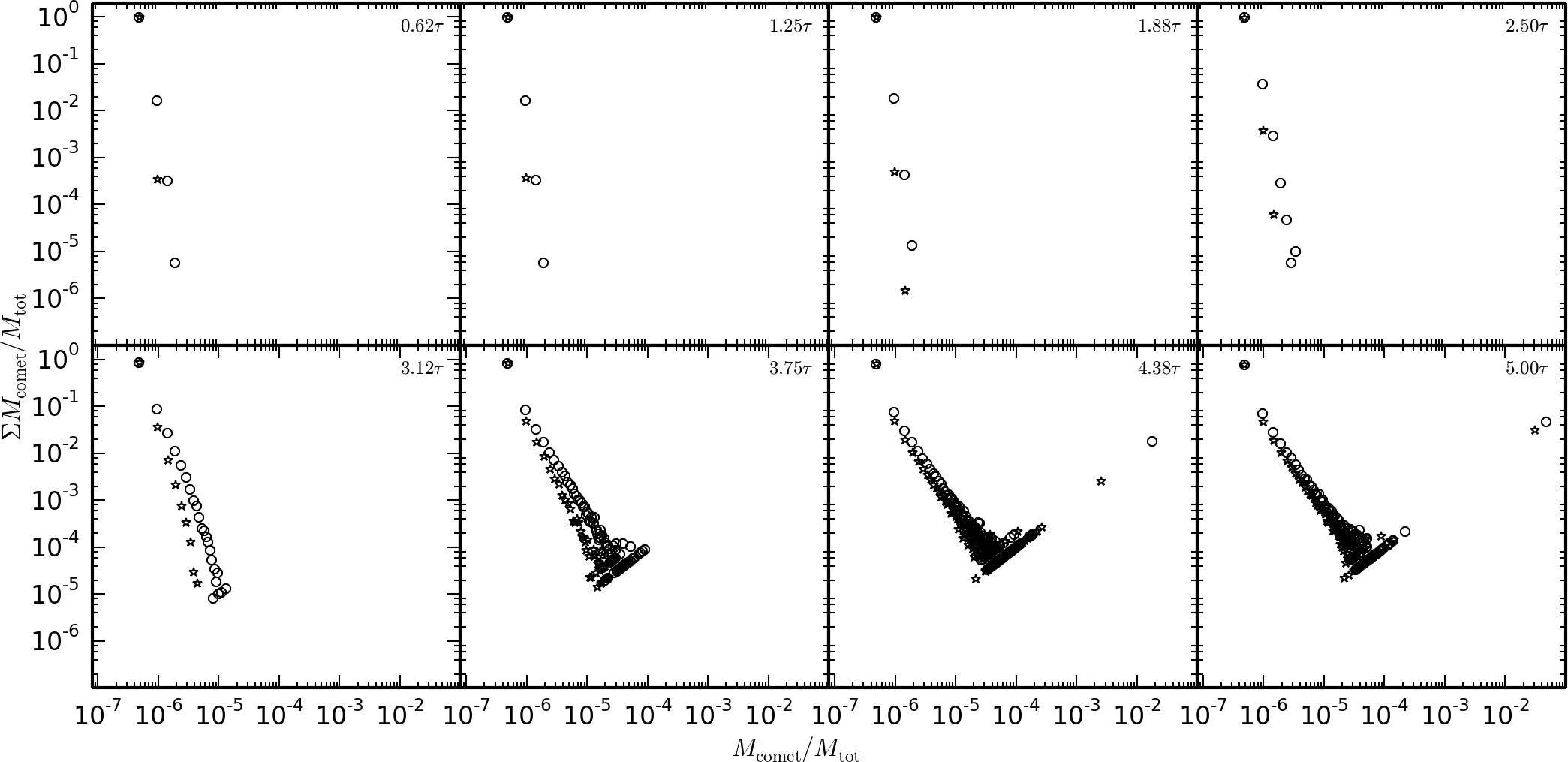}
  \caption{Snapshot ``comet'' mass-distribution summation sequence of OPGs simulations. Reference with
    $N_\name{SM} = 128^3$: circle; sum of 4 ``comet'' sizes per star: $100^3$.}
  \label{fGall3}
\end{figure*}

Figure \ref{fGall} shows the ``comet'' mass distribution of all OPGs simulations. For small ``comets'', the percentage
for a same number of super-molecules is the same for all $N_\name{SM}$. For example, at $t=5\tau$, all simulations have
a fraction of $\sim 10^{-1}$ for comets consisting of two super-molecules.  These small clumps should be called
multimers as mentioned below. But, with increasing $N_\name{SM}$, the mass of a comet consisting of two super-molecules
diminishes since $M_\name{2\,SM} = 2/N_\name{SM}$.

On the other hand, for large ``comets'' and the ``planetoid'', the mass is invariant to $N_\name{SM}$. For example, the
``planetoid'' at $t = 5\tau$ has the same mass of $M_\name{planetoid} \approx 5\cdot 10^{-2} M_\name{tot}$ for all
$N_\name{SM}$.

The fact that the mass distributions are shifted to the left for larger $N_\name{SM}$ is misleading, as shown in Fig.
\ref{fGall3}. Only the two largest simulations are shown, the simulation with $N_\name{SM} = 128^3$ is shown normally,
but for the simulation with $N_\name{SM} = 160^3$ is downscaled to $160^3/2 \approx 128^3$, always two ``comet'' sizes
are added together. As can be seen, when comparing the two simulation using the same scaling, the two mass distributions
are in fact very similar.

It is interesting to look at the turning point, defined as the point where the mass sum of the biggest ``comets'' is
larger than the mass sum of smaller comets. This point is interesting as it is the first indicator of the creation of a
aggregate of molecules that start to be influenced by gravity. Figure \ref{fturn} shows the time and the mass of the
turning point. One can see than the time of appearance is independent of $N_\name{SM}$. The ``comet'' mass, on the other
hand, clearly follows a power law with:
\begin{equation}
  \label{ePN}
  {M_\name{comet}\over M_\name{tot}}  \approx 10 N_\name{SM}^{\xi_\name{N}} \ ,
\end{equation}
with $\xi_\name{N} \approx -1$.  Since $N_\name{SM} = M_\name{tot}/M_\name{SM}$, at turning point the number of
super-molecules is always of order of 10.  In other words, in critical conditions where the attractive molecular force
adds up to gravity, the tendency to make a larger condensed body mass fraction already starts at about ten molecules.

\subsubsection{Extrapolation to physical scale}

\begin{figure}[t] 
  \resizebox{\hsize}{!}{\includegraphics{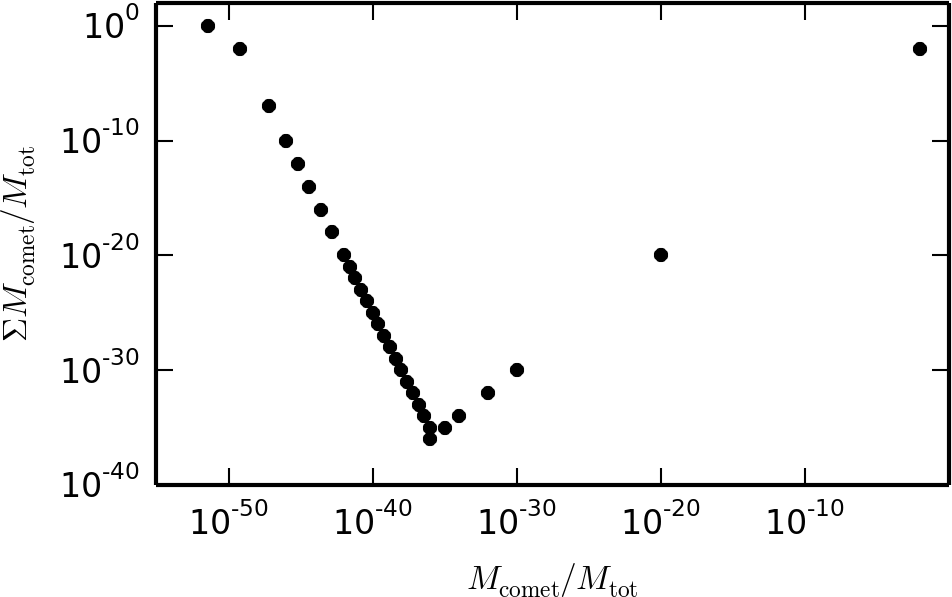}}
  \caption{Extrapolation of the ``comet'' mass distribution for a real H$_2$ molecular fluid at $t=5\tau$.}
  \label{fextra}
\end{figure}

Having studied the scaling behaviour of the simulations, we can attempt to extrapolate some properties for a fluid
consisting of real molecules instead of super-molecules. Considering H$_2$ molecules, the total mass of the OPGs
simulations is $1.4\,M_\oplus$, the box length equal to $2.3\cdot 10^{-3}\,\name{AU}$ and
$\tau = 1.5\cdot 10^{-2}\,\name{years}$.

Thus, in realistic conditions there would be a total amount of $2.5\cdot 10^{51}$ H$_2$ molecules, the number of
molecules per super-molecule $\eta$ varying from $2.0\cdot 10^{46}$ for $N_\name{SM} = 50^3$ to $3.2\cdot 10^{44}$ for
$N_\name{SM} = 160^3$. So, even for the largest simulations, there is a difference in number of particles of more than
$40$ orders of magnitude with real situations. For that reason, the extrapolated data has to be treated with caution.
As long as no other physics enters in the interval of scales, this is however not extraordinary in regards to common
practices in simulation works, such as simulating stars with SPH particles. While a single SPH particle can at best
represent a mass fraction of $10^{-6}-10^{-10}$ of the total, an SPH particle is supposed to behave as the smallest mass
element in local thermal equilibrium, say a few 1000 protons over the stellar mass: $\sim 10^{-55}$, which lies
therefore well over $10^{40}$ times the SPH simulation resolving capacity.

As can be observed in Fig. \ref{fGc}, the fraction of bound molecules does not change when increasing the number of
particles and will remain unchanged for a fluid consisting of molecules. The same is true for the size of the resulting
``planetoid'', as can be seen in Fig. \ref{fGall}. Therefore, we can assume that after $5\tau$, the system will consist
of $\sim 80\%$ unbound molecules and a ``planetoid'' with a mass of $\sim 0.07\,M_\oplus$.

At $t = 5\tau$, there is a fraction of $10^{-1}$ of two molecules aggregates i.e. dimers with a mass of
$6.6\cdot 10^{-27}\,\unit{kg}$.  These multi-molecules clumps should be called ``multimers'' instead of ``comets''.  As
mentioned above the turning point occurs at $\sim 2.4\tau$ consistently for approximately $10$ molecules, so its mass is
about $ 20 \,m_\name{H} = 3.3\cdot 10^{-26}\,\name{kg}$.

Using the above values and the two power laws of Equ. (\ref{ePC} -- \ref{ePP}), Fig.~\ref{fextra} shows a schematic
diagram showing the mass distribution at $t=5\tau$.  While the mass fraction of the turning point multimers is tiny, the
power-law distribution rapidly increases the mass in larger grain- and comet-sized bodies until they become a planetoid
mass.

\subsection{Phase transition} 
\label{ssTPS}

\begin{figure*}[t]
  \centering
  \includegraphics[width=18cm]{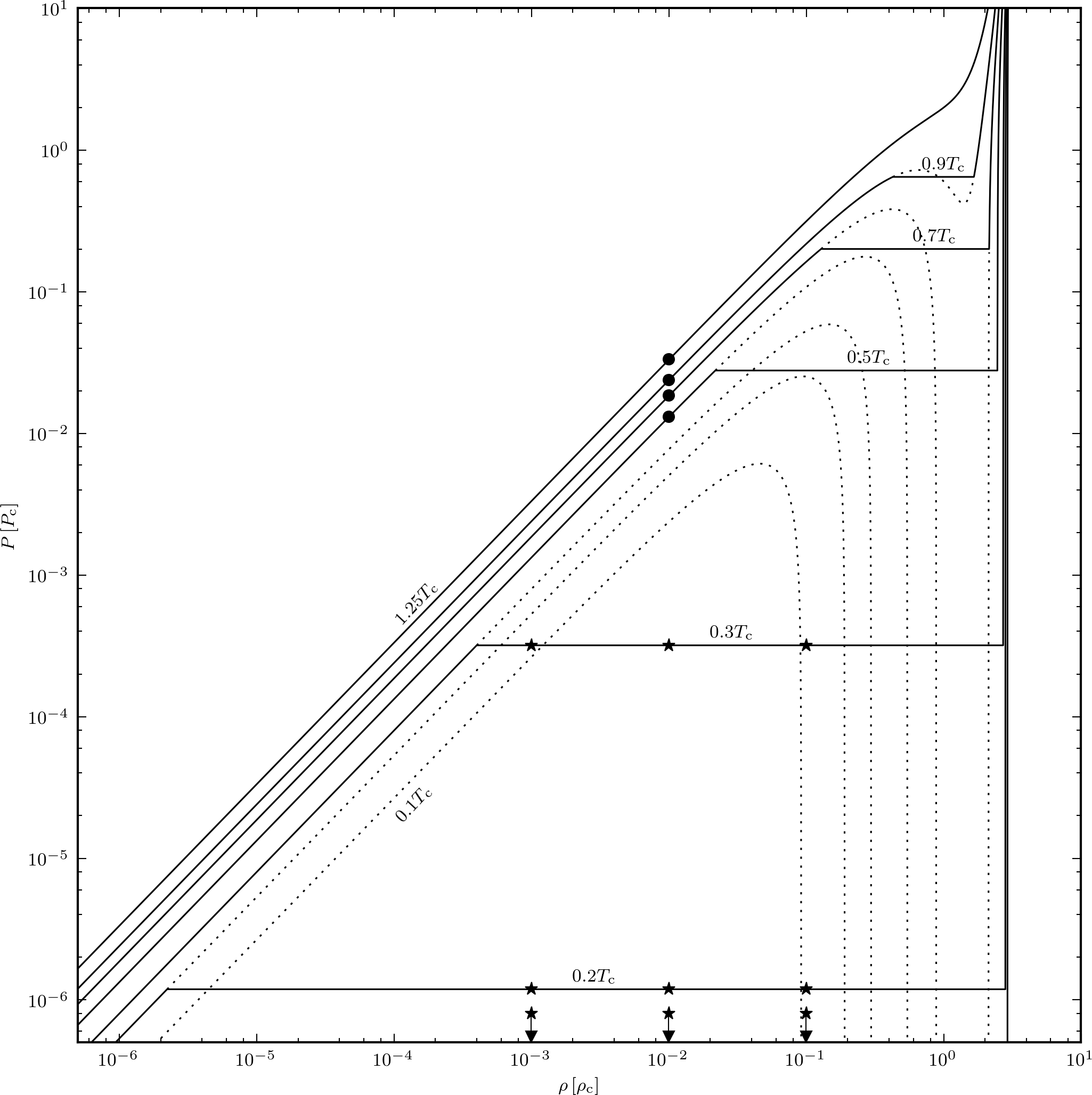}
  \caption{Position of the LJ simulations in a corresponding van der Waals phase diagram. Solid line: Maxwell construct
    for the van der Waals EOS; dotted line: van der Waals EOS.  Star: fluid presenting a phase transition; bullet:
    one-phase fluid.}
  \label{fvdwTPh}
\end{figure*}

We study several physical conditions close to a phase transition. Table \ref{tTP} summarizes their properties. Three
different densities, $10^{-1}$, $10^{-2}$ and $10^{-3}\,n_\name{cr}$, and three different temperatures, $0.1$, $0.2$ and
$0.3\,T_\name{cr}$, are chosen. As the one-phase fluid studied in Sect.~\ref{ssOPS} also has a density of
$10^{-2}\,n_\name{cr}$, three additional temperatures, $0.5$, $0.7$ and $0.9\,T_\name{cr}$, were used for this density
to make a link between the one-phase fluid and the fluids studied in this section.

Looking at the phase transition diagram (Fig.~\ref{fvdwTPh}), all fluids with $T \leq 0.3 T_\name{c}$ are on the Maxwell
line. Those with $\rho \leq 10^{-2}\rho_\name{c}$ are on the $(\partial P / \partial \rho)_\name{s} >0$ part of the van
der Waals phase diagram, which corresponds to metastable states on the verge of a deep phase transition. The fluids with
$\rho = 10^{-1}\rho_\name{c}$ are on the $(\partial P / \partial \rho)_\name{s} \leq 0$ part of the van der Waals phase
diagram, which corresponds to unstable states in a phase transition. The fluids with $T \geq 0.5 T_\name{c}$ are still
in the stable regime.

Three different cases are studied: without gravity, strongly self-gravitating above the Jeans criterion, and weakly
self-gravitating below the Jeans criterion.

\subsubsection{Fluid without gravity}
\label{ssFwg}

\begin{figure}[t] 
  \resizebox{\hsize}{!}{\includegraphics{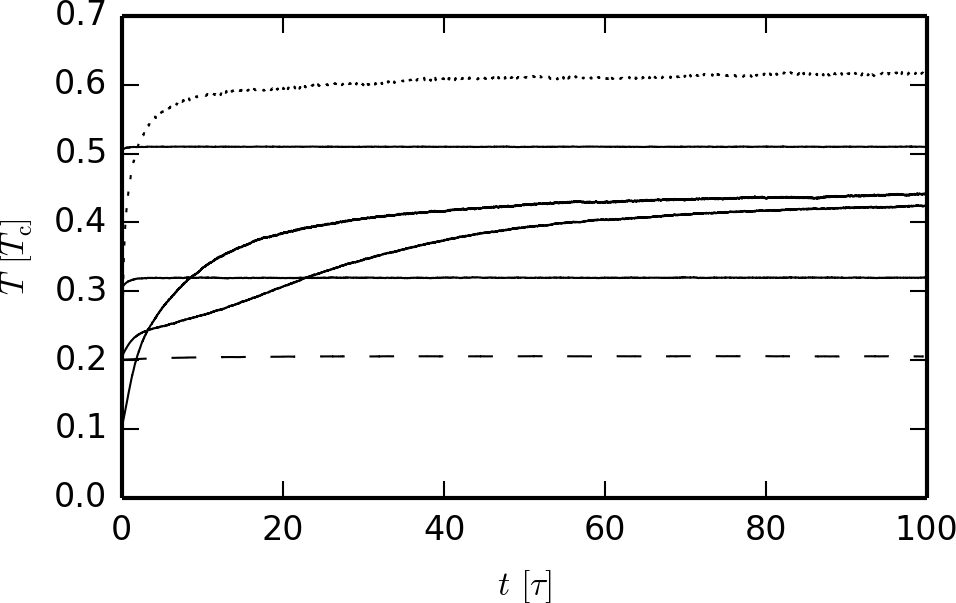}}
  \caption{Temperature of non-gravitating fluids as a function of time. Dotted line: $n = 10^{-1} n_\name{cr}$; solid
    line: $n = 10^{-2} n_\name{cr}$; dashed line: $n = 10^{-3} n_\name{cr}$. All simulations with $N_\name{SM} = 50^3$.}
  \label{fTP0}
\end{figure}

\begin{figure}[t] 
  \resizebox{\hsize}{!}{\includegraphics{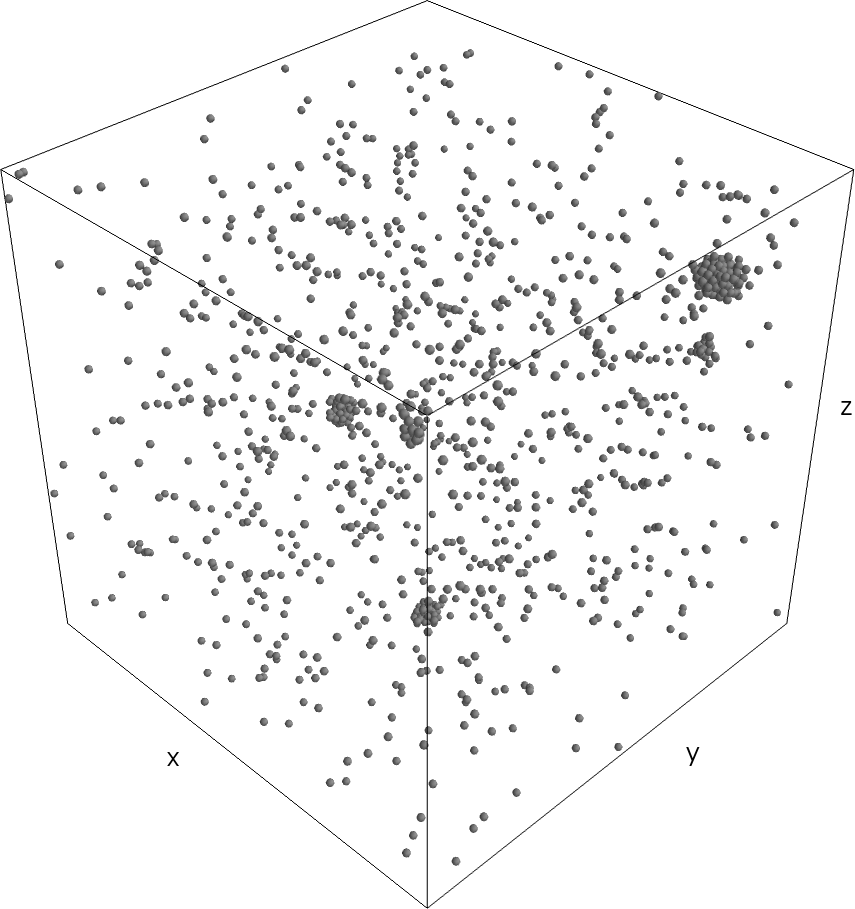}}
  \caption{Close-up 3D-view of a medium sized ``comet'' and smaller aggregates.}
  \label{fclump}
\end{figure}

\begin{figure}[t] 
  \resizebox{\hsize}{!}{\includegraphics{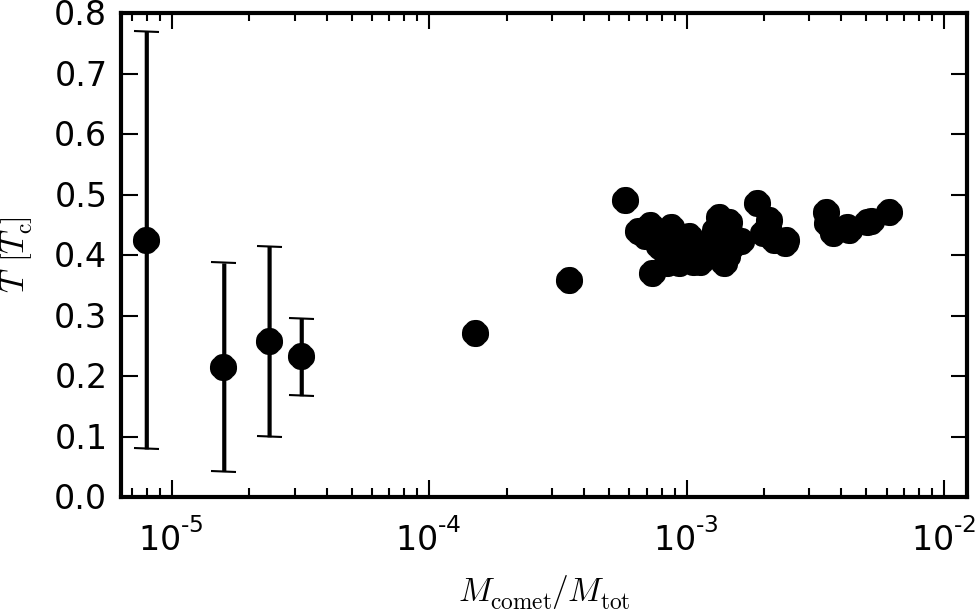}}
  \caption[Temperature distribution of unbound molecules and ``comets'' as a function of ``comet'' size of PT2-2 without
  gravity and $N_\name{SM} = 50^3$ at $t = 100\tau$.] {Temperature distribution of unbound molecules and ``comets'' as a
  	function of ``comet'' size of PT2-2$^a$  without gravity and $N_\name{SM} = 50^3$ at $t = 100\tau$. \\ $^{(a)}$ See
  	Table \ref{tTP}}
  \label{fTPT} 
\end{figure}

To study the evolution of fluids presenting a phase transition without gravity, the $x_\name{GLJ}$ value does not have
to be considered and the number of super-molecules is set to $N_\name{SM} = 50^3$. Figure \ref{fTP0} shows the
temperature evolution of some of these fluids. In the beginning, the molecules merge into small ``comets'', which leads
to a decrease of the potential energy and therefore a rise of the temperature. Once the temperature is high enough, the
kinetic and potential energy reach a equilibrium and the fluid remains stable.

The number of ``comets'' formed is dependent on the density and inversely dependent on the temperature. The higher the
number of formed ``comets``, the longer it takes for the fluid to reach a stable regime.  For example, the very cold
fluids PT1-2 and PT2-2 only reach an asymptotic value around $t \approx 100\tau$.

\begin{figure*}[p] 
  \centering
  \includegraphics[width=18cm]{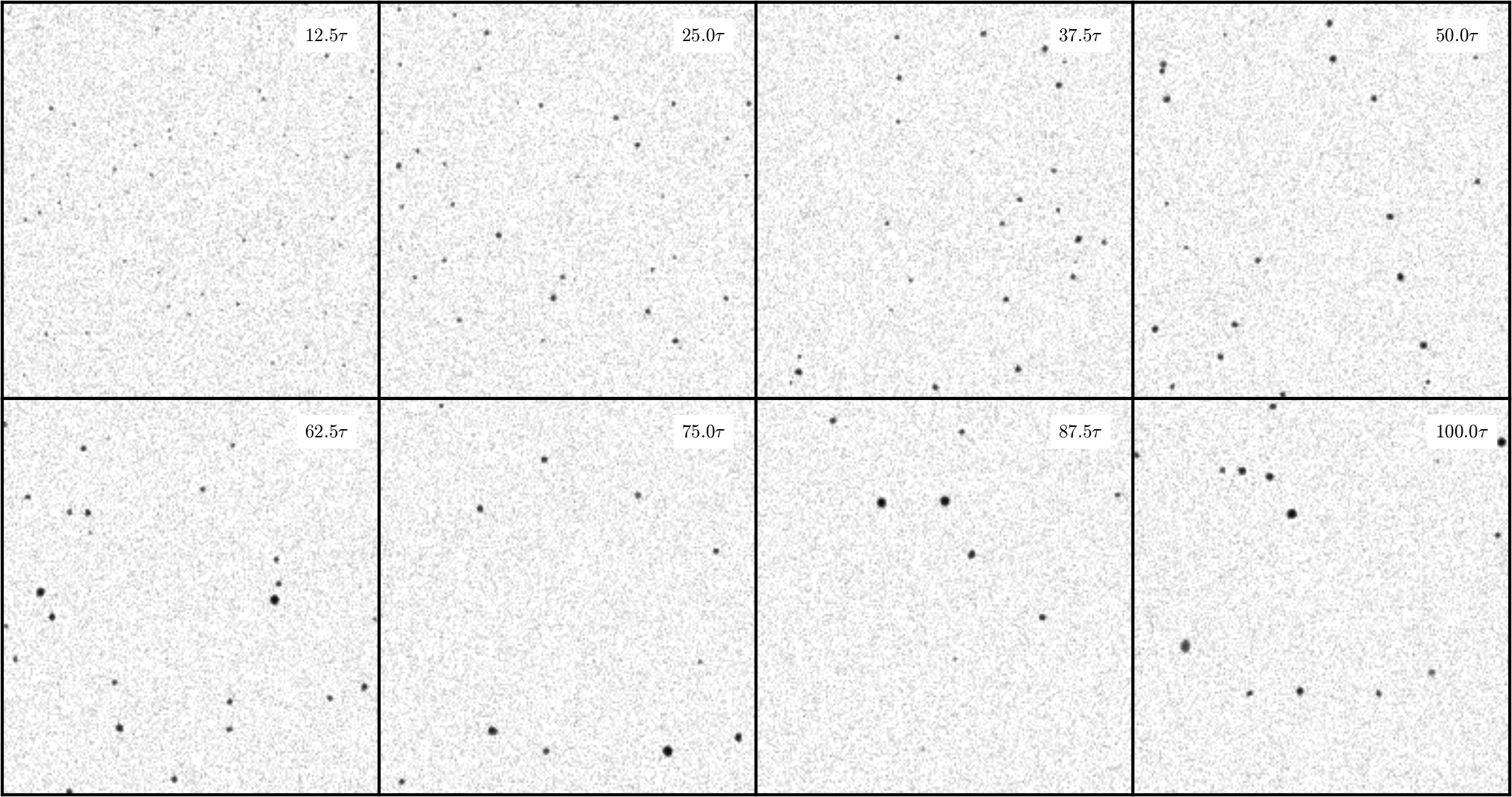}
  \caption{$0.2\times 1\times 1 \,L^3$ snapshots at $x=(0.5\pm0.1)L$ of simulation PT2-2 with $N_\name{SM} = 50^3$, and
    without gravity.}
  \label{fPT2-2s}

  \medskip

  \centering
  \includegraphics[width=18cm]{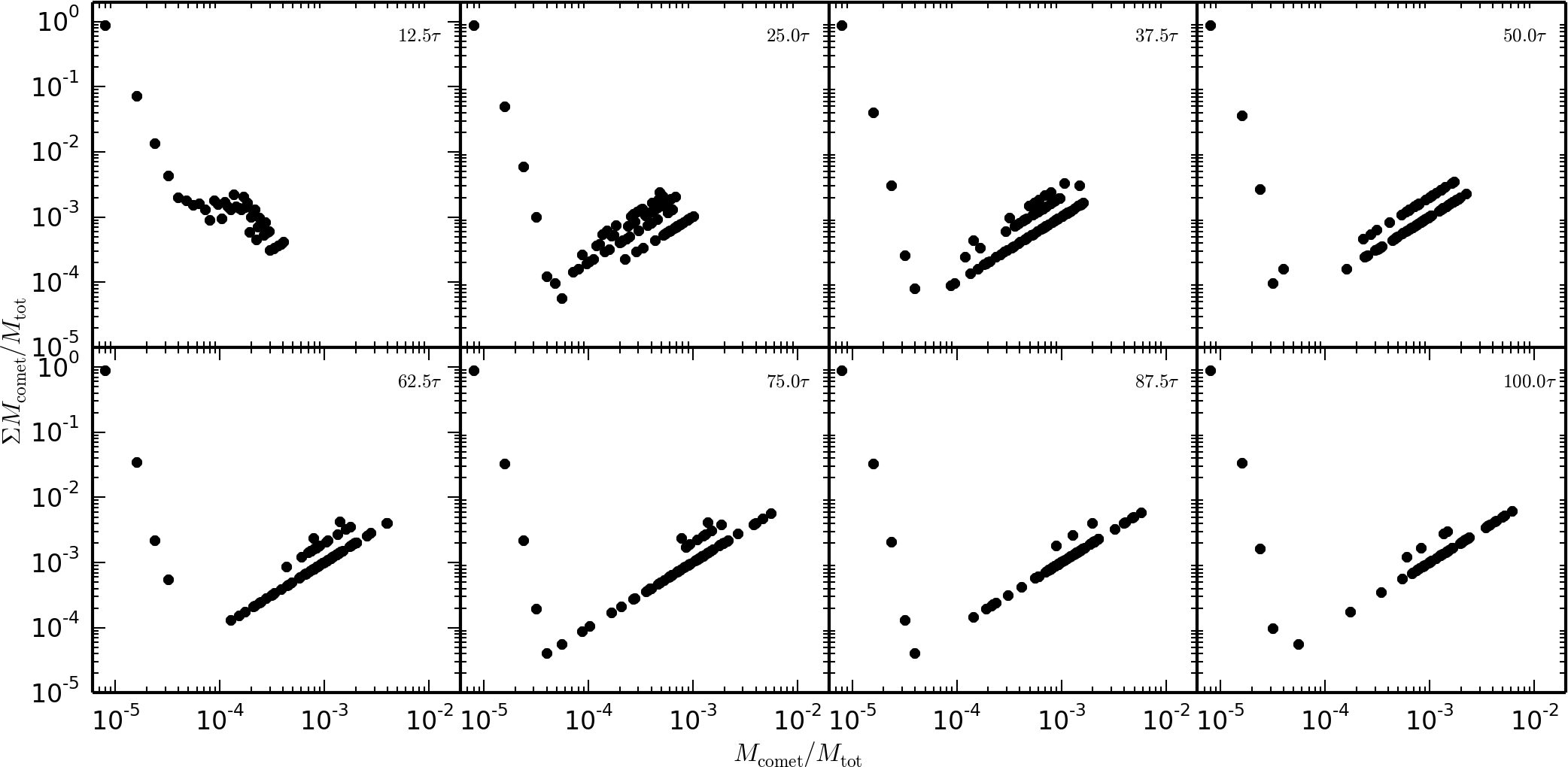}
  \caption{``Comet'' mass-distribution snapshots of simulation PT2-2 with $N_\name{SM} = 50^3$, and without gravity.}
  \label{fPT2-2h}
\end{figure*}

The ``comets'' remain at moderate size as can be seen in the snapshots (Fig.~\ref{fPT2-2s}) and in their mass
distribution (Fig.~\ref{fPT2-2h}): The number of super-molecules per ``comets'' is $< 10^{-2}\, N_\name{SM}$. The mass
distribution follows the power law of Equ. (\ref{ePC}), but at $\sim 10^{-4}$, the mass distribution rises again,
presenting a big number of ``comets'' between $10^{-4}$ to $10^{-2}$ super-molecules. Fig.~\ref{fclump} shows a close-up
3D view of a medium-sized ``comet'' and smaller aggregates.

Figure \ref{fTPT} shows the temperature distribution as a function of the ``comet'' mass. As in simulation OPGs, the
temperature for small bodies lies below average, but reaches average value for more massive bodies.

\subsubsection{Self-gravitating fluid above Jeans criterion}

\begin{figure}
  \resizebox{\hsize}{!}{\includegraphics{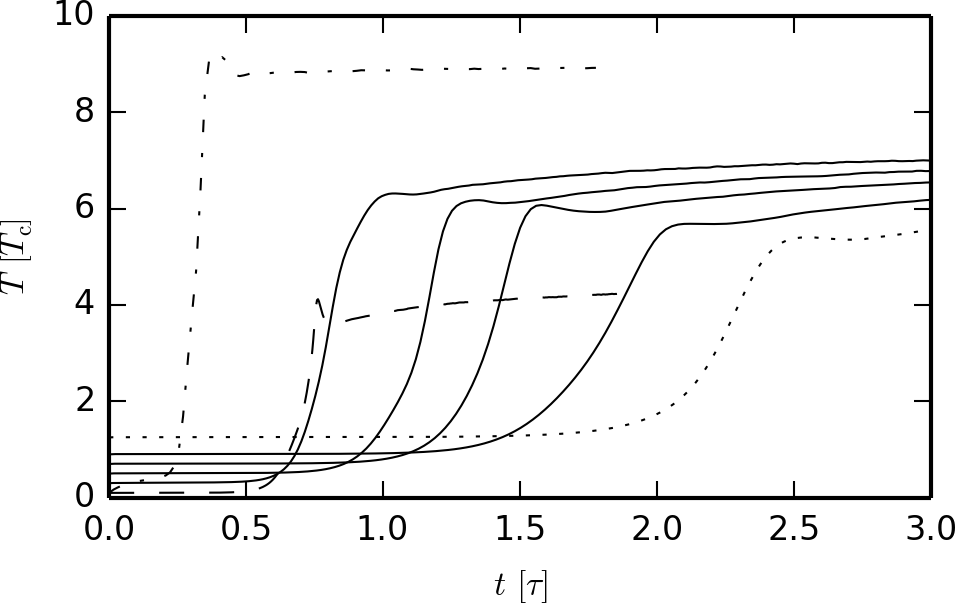}}
  \caption[Temperature of sufficiently self-gravitating fluids as a function of time. Dotted line: OPGs; dashed line:
  PT1-3; solid line (from left to right): PT3-2, PT5-2, PT7-2, PT9-2; dash-dotted line: PT1-1. All simulations with
  $N_\name{SM} = 100^3$.] {Temperature of sufficiently self-gravitating fluids as a function of time. Dotted line:
  	OPGs$^a$; dashed line: PT1-3$^a$; solid line (from left to right): PT3-2$^a$, PT5-2$^a$, PT7-2$^a$, PT9-2$^a$;
  	dash-dotted line: PT1-1$^a$. All simulations with $N_\name{SM} = 100^3$. \\ $^{(a)}$ See Table \ref{tTP}}
  \label{fTP}
\end{figure}

All sufficiently self-gravitating simulations use the same gravitational potential,
$G = 1.25\,G_\name{J}\left(T = 1.25\,T_\name{cr}, n = 0.01\,n_\name{cr}\right)$. As the temperatures and densities
differ, $G_\name{J}$ and therefore $\gamma_\name{J}$ is different for each simulation, but this parameter is always
$> 1$.

As $\gamma_\name{J} > 1$ for all simulated fluids, they are unstable if perturbed. The lower the initial temperature of
a fluid, the bigger is $\gamma_\name{J}$, which translates to a faster exponential growth. This can be seen in
Fig.~\ref{fTP} where the fluids with a density of $10^{-2}\,n_\name{cr}$ are starting to collapse one after another,
from the lowest to the highest temperature.

\begin{figure*}[p] 
  \centering
  \includegraphics[width=18cm]{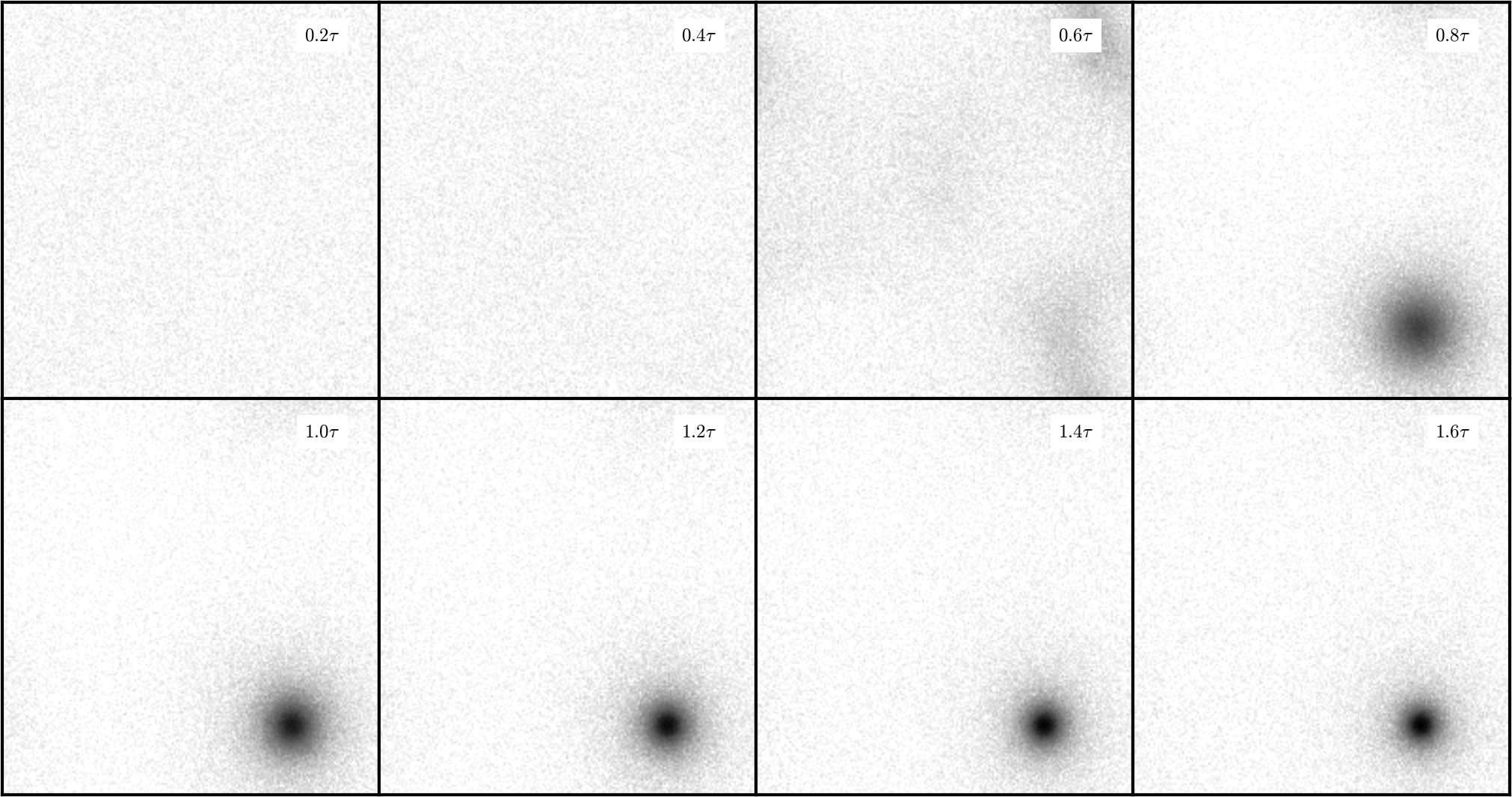}
  \caption{$0.025 \times 1\times 1 \,L^3$ snapshots at $x=(0.5\pm0.0125)L$ of sufficiently self-gravitating simulation
    PT1-3 with $N_\name{SM} = 100^3$.}
  \label{fPT1-3s}

  \medskip

  \centering
  \includegraphics[width=18cm]{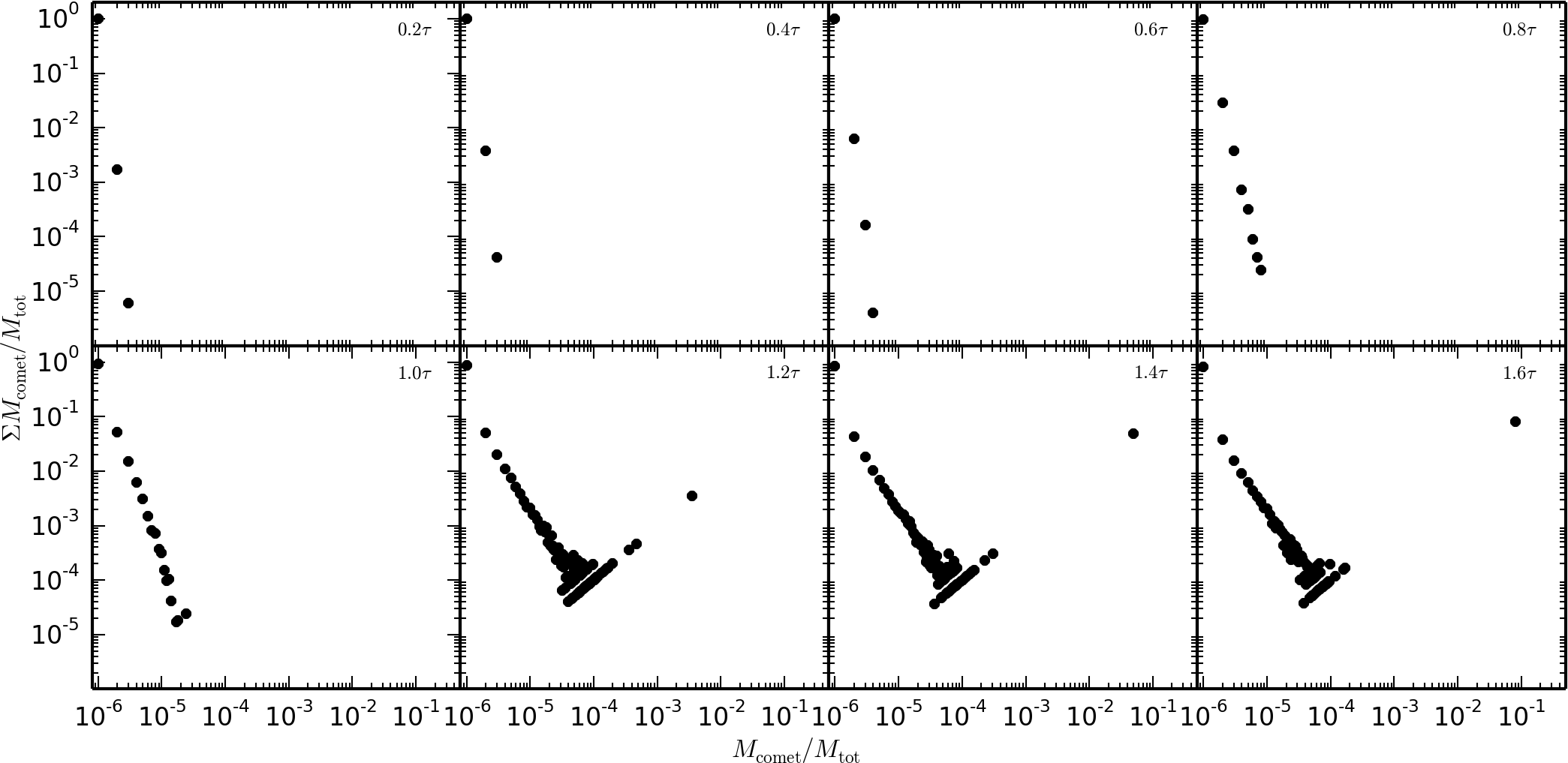}
  \caption{``Comet'' mass-distribution snapshots of sufficiently self-gravitating simulation PT1-3 with
    $N_\name{SM} = 100^3$.}
  \label{fPT1-3h}
\end{figure*}

The evolution of the fluids is identical to the fluid OPGs: an exponential rise of the temperature until it reaches a
temperature break, after which the temperature rises only slowly and a single ``gaseous planetoid'' is formed. Figures
\ref{fPT1-3s} and \ref{fPT1-3h} show snapshots and ``comet'' mass distribution, again very similar to the fluid OPGs.
First, there is a formation of small ``comets'' following the power law of Equ. (\ref{ePC}), and then the ``comets'' are
massive enough to attract each other and merge into one big spherical ``planetoid'' ($t> 1\tau$), following the power
law of Equ. (\ref{ePP}). The power-law index $\xi_\name{c}$ also evolves in a similar fashion as in simulation OPGs,
starting out very steep and reaching a final value of $\gtrsim -2$.

\subsubsection{Self-gravitating fluid below Jeans criterion}

\begin{figure}[t] 
  \resizebox{\hsize}{!}{\includegraphics{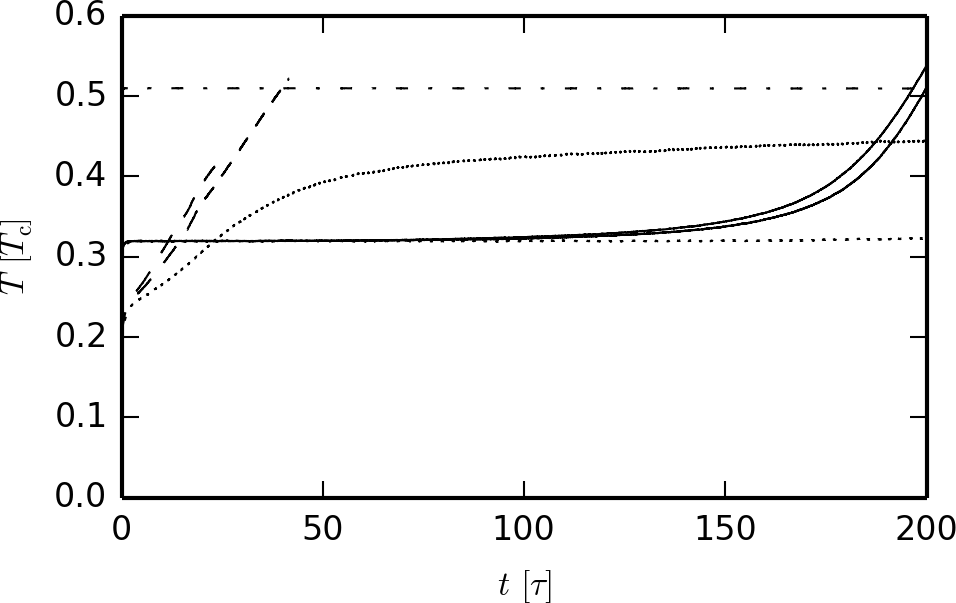}}
  \caption{Temperature of weakly self-gravitating fluids as a function of time. Dotted line: PT2-2 and PT3-2, with
    $N_\name{SM} = 50^3$ and without gravity; dashed line: PT2-2, $\gamma_\name{J} = 0.5$ and $N_\name{SM} = 80^3$ and
    $100^3$; solid line PT3-2, $\gamma_\name{J} = 0.5$ and $N_\name{SM} = 80^3$ and $100^3$; dash-dotted line PT5-2,
    $\gamma_\name{J} = 0.5$ and $N_\name{SM} = 100^3$.}
  \label{fl}
\end{figure}

\begin{figure}[t] 
  \resizebox{\hsize}{!}{\includegraphics{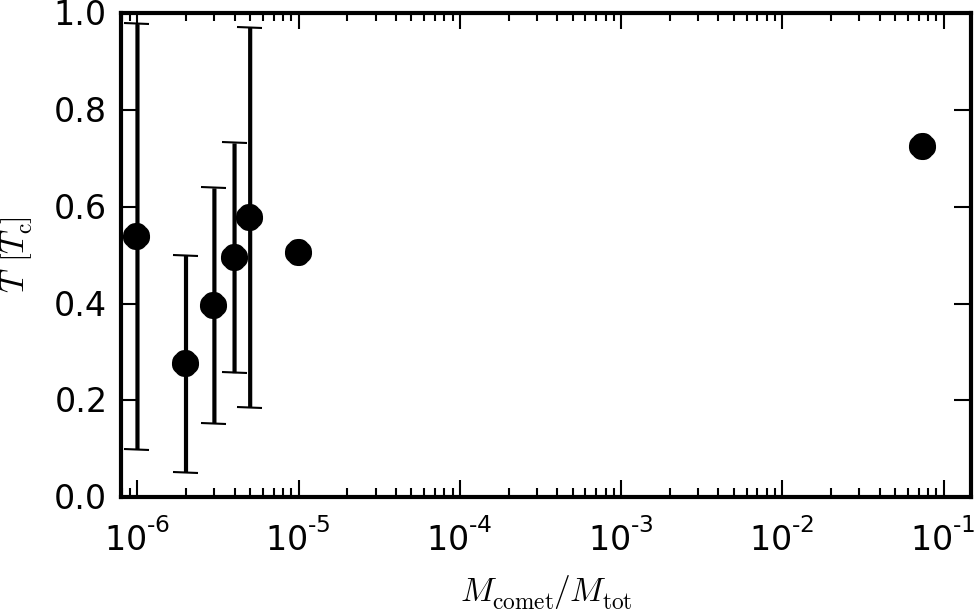}}
  \caption{Temperature distribution of unbound molecules and ``comets'' as a function of ``comet'' size of PT3-2 with
    $\gamma_\name{J} = 0.5$ and $N_\name{SM} = 100^3$ at $t = 200\tau$.}
  \label{fTPlT}
\end{figure}

To compare weakly self-gravitating fluids in a phase transition (on the Maxwell line) with out-of-phase-transition
fluids (below the Maxwell line), $\gamma_\name{J}$ is set to $0.5$ for all fluids. This means that the individual $G$
value is different for each simulation.

As shown in Sect. \ref{ssFwg}, the perturbation of the weakly self-gravitating one-phase fluid OPGw does not create any
effect if $\gamma_\name{J} < 1$. Figure \ref{fl} shows the temperature as a function of time of the self-gravitating
fluids PT5-2, PT3-2, and PT2-2, and compares them to the non-gravitating fluids. The out-of-phase-transition fluid PT5-2
remains unchanged and does not differ from the non-gravitating fluid, identical to OPGw.

The temperatures of the fluids in a phase transition PT3-2 and PT2-2 are exponentially growing and differ clearly from
the non-gravitating ones. In the beginning, the LJ forces dominate the gravitational forces and the fluids behave like
non-gravitating fluids. PT2-2 reacts much faster than PT3-2, thanks to its low initial temperature and fluctuations due
to the Maxwellian velocity distribution,and forms many small- and medium-sized ``comets'' (see Sect. \ref{ssFwg}). Once
these ``comets'' are formed, their mass is high enough to attract each other with gravity, forming a single ``rocky
planetoid''.

\begin{figure*}[p] 
  \centering
  \includegraphics[width=18cm]{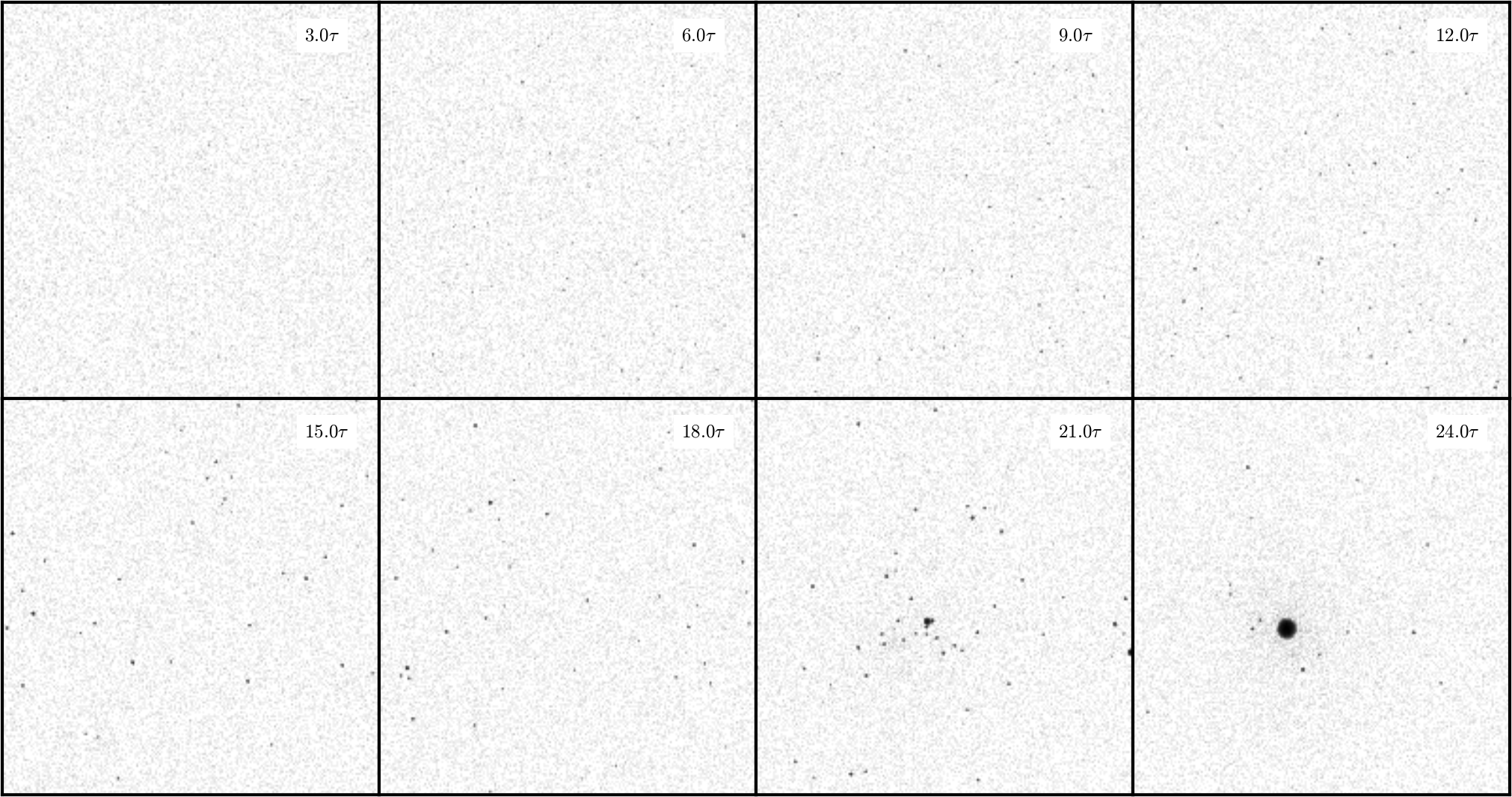}
  \caption{$0.025 \times 1\times 1 \,L^3$ snapshots at $x=(0.5\pm0.0125)L$ of weakly self-gravitating simulation PT2-2
    with $N_\name{SM} = 100^3$.}
  \label{fPT1-2s}

  \medskip

  \centering
  \includegraphics[width=18cm]{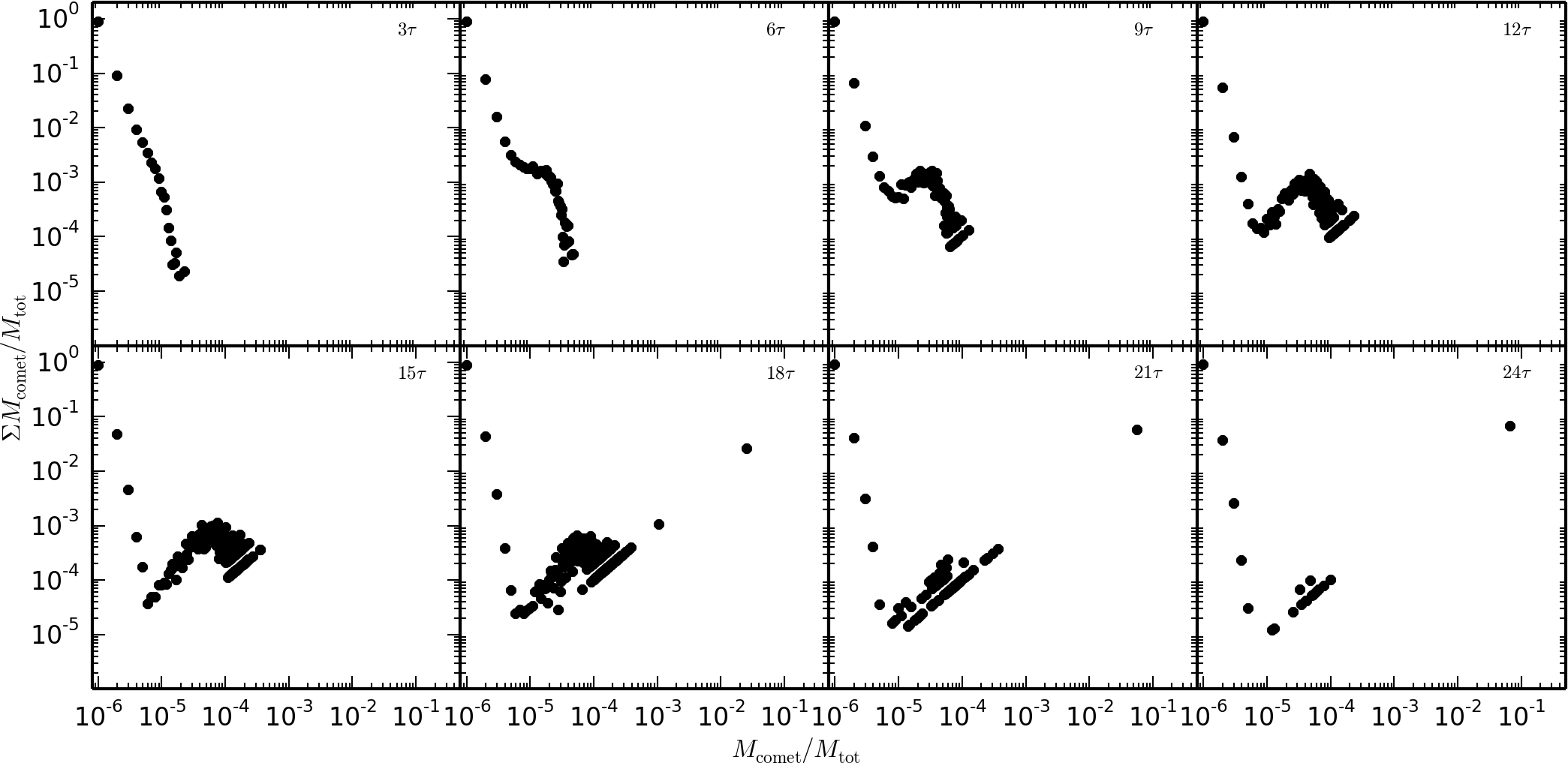}
  \caption{``Comet'' mass distribution of weakly self-gravitating simulation PT2-2 with $N_\name{SM} = 100^3$.}
  \label{fPT1-2}
\end{figure*}

This can be seen in the snapshots (Fig. \ref{fPT1-2s}) and ``comet'' mass distribution (Fig. \ref{fPT1-2}). During the
first $15\tau$, the fluid is identical to the low-temperature, non-gravitating fluid (compare to
Fig. \ref{fPT2-2h}). Only then does gravity start to play a role, one can see the formation of a ``planetoid'' and how
it swallows the small- and medium-sized ``comets'' during its growth.

\begin{figure*}[p] 
  \centering
  \includegraphics[width=18cm]{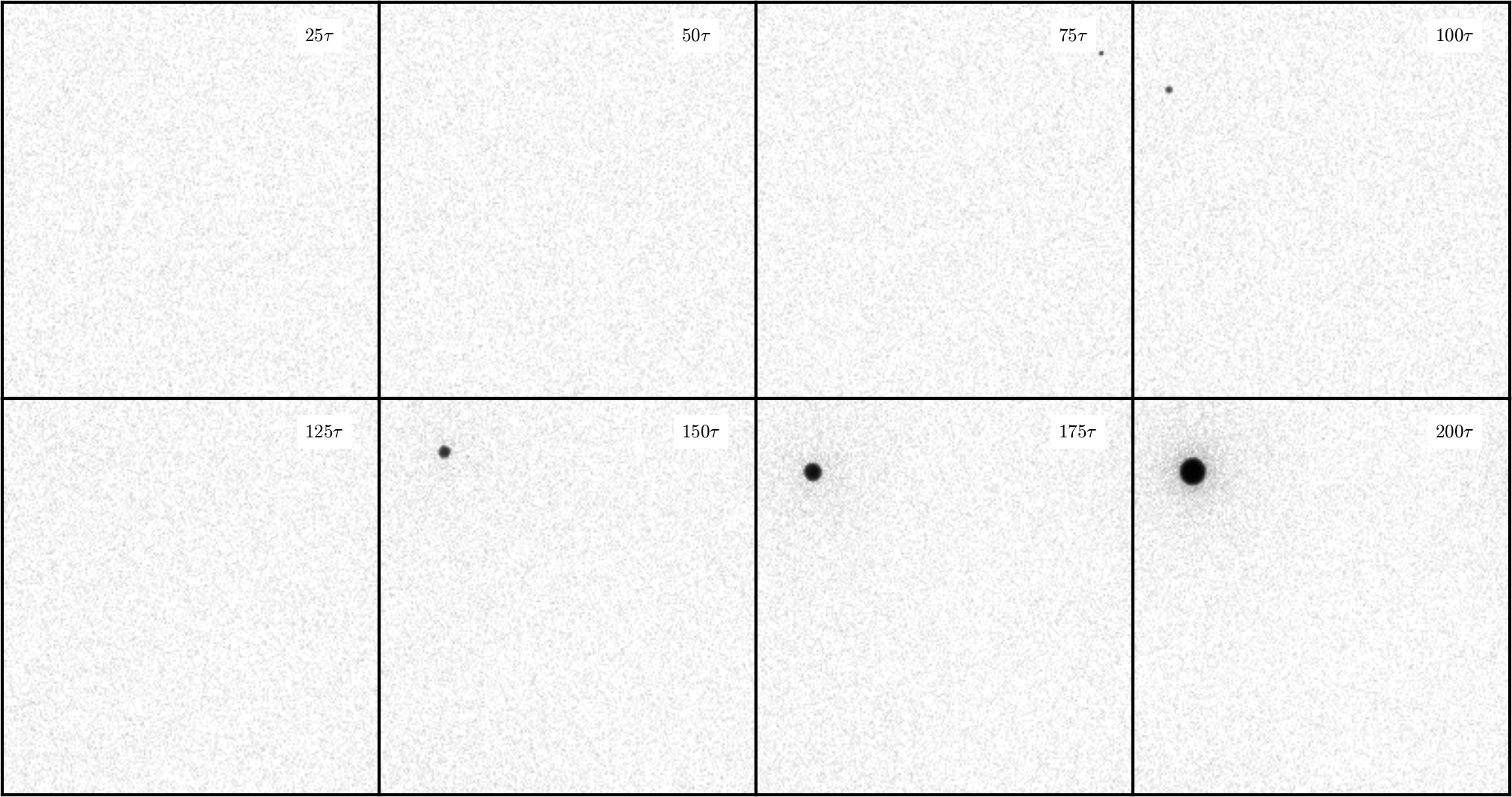}
  \caption{$0.025 \times 1\times 1 \,L^3$ snapshots at $x=(0.5\pm0.0125)L$ of weakly self-gravitating simulation PT3-2
    with $N_\name{SM} = 100^3$.}
  \label{fPT3-2s}

  \medskip

  \centering
  \includegraphics[width=18cm]{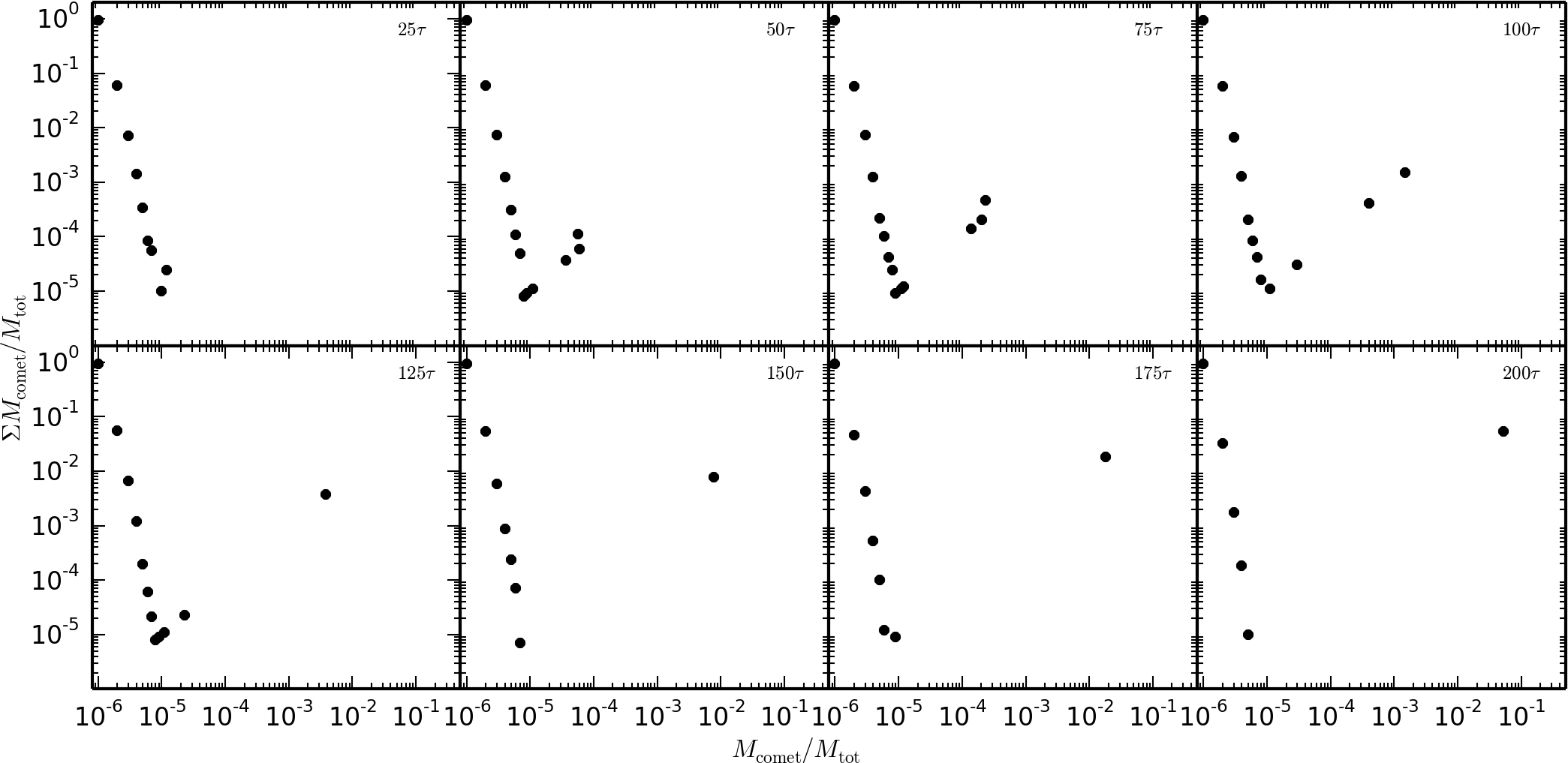}
  \caption{``Comet'' mass distribution of weakly self-gravitating simulation PT3-2 with $N_\name{SM} = 100^3$.}
  \label{fPT3-2}
\end{figure*}

With its higher temperature, PT3-2 forms almost no small- or medium-sized ``comets'', which can be seen in the snapshots
(Fig. \ref{fPT3-2s}) and ``comet'' mass distribution (Fig. \ref{fPT3-2}) for the first $\approx 50 \tau$. Only after
$50 \tau$ does gravity show its effect with the appearance of a medium-sized ``comet'', which is too massive to fit into
the power law $\xi_1$ (see Fig. \ref{fPT3-2s}, $t = 50\tau$). From then on, this ``comet'' attracts super-molecules and
thus grows in size until at $t \approx 125\tau$, a single big ``rocky planetoid'' is formed.

Figure \ref{fTPlT} shows the temperature distribution as a function of the ``comet'' mass of PT3-2. The temperature of
the ``comets'' containing few super-molecules lies below the average, but the ``planetoids'' temperature is the same as
the average.

PT2-2, PT3-2 and PT5-2 all scale very well. The two fluids in a phase transition, PT2-2 and PT3-2, have an almost
identical timescale for the exponential growth for $N_\name{SM} = 80^3$ and $100^3$. The slight difference is due to the
initial random seed. As $x_\name{GLJ}$ depends on $\gamma_\name{J}$ and the temperature, both much lower than for OPGs,
its value is clearly above the cut-off radius, which is $7.63$ and $8.34$ for PT2-2 and $7.03$ and $7.69$ for PT3-2.

PT5-2 shows no difference for any number of super-molecules, with $N_\name{SM}$ ranging from $50^3$ to $100^3$.

\subsection{``Planetoid'' densities}
\begin{figure}[t] 
  \resizebox{\hsize}{!}{\includegraphics{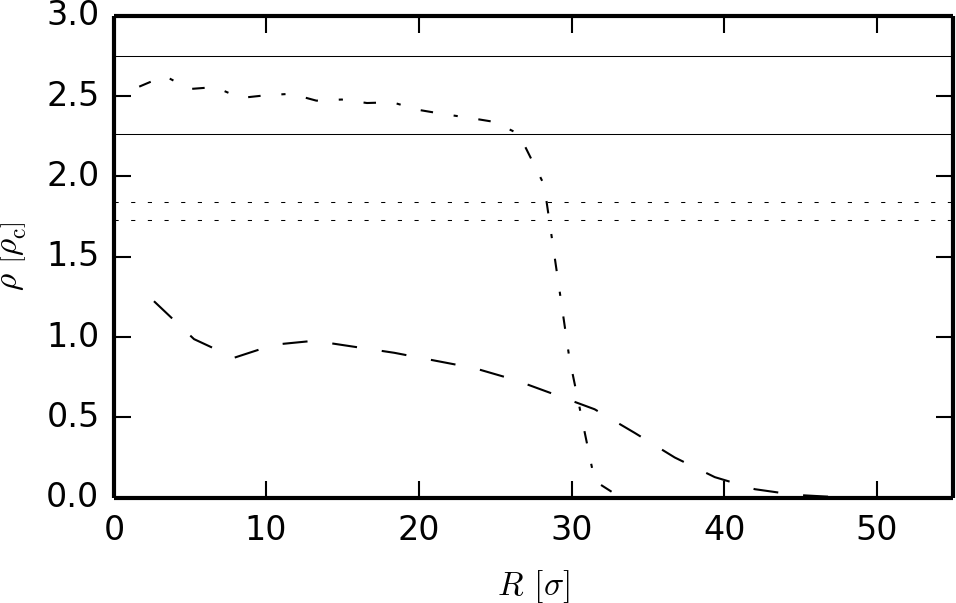}}
  \caption{Density of ``planetoid'' as a function of the radius. Dashed line: OPGs with $N_\name{SM} = 100^3$ at
    $t=4\tau$, the ``gaseous planetoid'' consists of 41559 super-molecules ($0.042\,M_\name{tot}$). Dash-dotted line:
    PT3-2, with $\gamma_\name{J} = 0.5$ and $N_\name{SM} = 100^3$ at $t = 200\tau$; the ``rocky planetoid' consists of
    74208 super-molecules ($0.074\,M_\name{tot}$). Solid lines: hydrogen, higher value: solid; lower value:
    liquid. Dotted line: $\rho_\mathrm{f}$ of Lagrange-Jacobi identity, higher value: SC; lower value:
    HCP/FCC.} \label{fPlanetD}
\end{figure}

Figure \ref{fPlanetD} shows the densities of the ``planetoids'' as a function the radius of the simulations OPGs and
PT3-2 with $\gamma = 0.5\gamma_\name{J}$, and compares them to hydrogen laboratory data and values found in the
Lagrange-Jacobi identity (Sect. \ref{ssV}).

The density of the OPGs ``gaseous planetoid'' is below that of a conventional solid or liquid.  As their temperature is
above the critical value ($T> 6\,T_\name{cr}$), super-molecules are not able to condense without the aid of gravity, and
because of their high kinetic energy, are vibrating at much higher amplitudes. This leads to more space between the
super-molecules and therefore a lower density compared to super-molecules below the critical temperature.

The density of the ``gaseous planetoid'' drops with increasing radius and hence does not qualify as a homogeneous
sphere, but its average density is below the $\rho_\mathrm{f}$ value of the Lagrange-Jacobi identity (Equ. \ref{erho0}),
in accordance with Tab. \ref{tLJ} and Fig. \ref{fvirial-ratios} for ``gaseous planetoids''.

The density of the PT3-2 ``rocky planetoid'' lies between the liquid and solid phase. There is no continuous density
drop as for the OPGs simulation, the density remains stable up to almost the outer radius and the body can be
approximated as a homogeneous sphere. Its density lies above $\rho_\mathrm{f}$ in accordance with Tab. \ref{tLJ} and
Fig. \ref{fvirial-ratios} for ``rocky planetoids''.

\section{Conclusions} 
\label{sC} 

We used analytic methods (Sect. 2) and computer simulations (Sects. 3-4) to study substellar fragmentation of fluids
presenting a phase transition. The motivating astrophysical context are molecular clouds where H$_2$ forms the bulk of
the gravitating mass and is not very far from condensation conditions.

\subsection{Analytic results}
The study of the virial theorem, using the gravitational and the Lennard-Jones potential energies, has shown that there
is a maximum temperature below which a fluid can fragment at arbitrary small masses. This temperature is an order of
magnitude above the critical temperature. This shows, granted the right circumstances, that ``comets'' can form at a
temperatures an order of magnitude larger than the critical temperature. In the case of H$_2$, this maximum temperature
lies in the range of $400$ -- $600\,\unit{K}$, depending on the solid crystalline structure.

A van der Waals fluid can be in three different states: gaseous, solid/liquid, or in a phase transition where the two
phases coexist. The latter is defined as lying on the line of the Maxwell construct, where
$(\partial P/\partial\rho)_\name{s} = 0$. A fluid is gravitationally unstable if an introduced perturbation has a
wavelength above a certain value, which depends on $(\partial P/\partial\rho)_\name{s}$.  Since this quantity vanishes
for a fluid presenting a phase transition, such a fluid is also gravitationally unstable.

\subsection{Simulation results}
We performed simulations using the state-of-the-art molecular dynamics simulator LAMMPS. We used super-molecules to
simulate gravitational and molecular forces together with a computationally tractable number of particles.

\subsubsection{Super-molecules}
We tested the super-molecule concept thoroughly. We achieved good scaling for non-gravitational effects if the number of
super-molecules is large enough ($\gtrsim 10^5$). The sticking point is to ensure that the molecular LJ forces are
dominating the gravitational forces in close-range interactions. This is achieved by setting the number of
super-molecules to be large enough so that the distance at which the two forces are equal is above the cut-off radius
$4\,\sigma_\name{SM}$.  For an H$_2$ fluid this sets a maximum super-molecule mass equal to
$5.7 \cdot 10^{-6}\,M_\oplus$, limiting the total mass that can be simulated by the available computing power.

In principle, the super-molecule concept should not perfectly reproduce the time dependence of diffusive properties, as
bigger particles introduce faster relaxation.  But our experiments using different resolutions spanning several orders
of magnitude did not reveal important modifications in the timings of major collapse and asymptotic evolution
state. This enables us to study the fragmentation of large bodies, which we call ``comets'' and ``planetoids''.

\subsubsection{One-phase fluid}
We applied a plane sinusoidal perturbation into a one-phase fluid with a temperature above the critical value. The
results reproduce the ideal-gas Jeans instability: no collapse is seen for conditions below the Jeans criterion, while
an exponential growth of the perturbation is observed for conditions above it.

As a result, the temperature rises, and small- and medium-sized ``comets'' form, some of which later merge into one big
``planetoid''. An interesting observation is the mass distribution of these ``comets'', which follows a power law for
the small- and medium-sized "comets", while the ``planetoid'' and the largest ``comets'' follow a different power law.

\subsubsection{Phase transition fluid}
We simulated fluids with temperatures below the critical value and close to the effective phase transition for three
different cases: without gravity, sufficiently, and weakly self-gravitating (above and below the ideal-gas Jeans
criterion).

Because of the Maxwellian velocity distribution fluctuations, the non-gravitating fluids form small- and medium-sized
``comets'' until the potential and kinetic energy reach an equilibrium; thereafter they remain statistically stable. In
the absence of any long-range force, no ``planetoid'' forms.

The self-gravitating fluids with a gravitational potential above the ideal-gas Jeans criterion do not react differently
to a plane sinusoidal perturbation from a one-phase fluid: the perturbation grows exponentially and its temperature
rises, and small- and medium-sized ``comets'' form, which ultimately merge into one ``planetoid''.

The analysis predicts that fluids presenting a phase transition are unstable even if the gravitational potential alone
is below the ideal-gas Jeans threshold. The performed simulations of weakly self-gravitating fluids in the phase
transition regime did reproduce the prediction. Because of the weak nature of the gravitational potential, the timescale
to see this reaction is two orders of magnitude longer than for sufficiently self-gravitating fluids, but, in the end, a
big central ``planetoid'' forms anyway. While the out-of-phase transition fluids do not amplify perturbations, those in
a phase transition indeed amplify them.

\bigskip 

This study is general enough to be relevant for many astrophysical situations. In the case of H$_2$ as main component,
it concerns situations where temperature may drop well below $10\,\name{K}$, such as in cold molecular gas in the outer
galactic disks, in expanding winds of planetary nebulae or supernovae, where cold substellar mass condensations are
known to form, or in protoplanetary disks. Other molecules besides H$_2$, such as CO, can condense even sooner, but as
their mass fraction is low they should not significantly perturb the gravitational balance, while He should not condense
at all and should keep a minimal amount of gaseous phase.\footnote{In \cite{safa_equation_2008} it was calculated that
  the mixture of He and H$_2$ at cosmic abundance does not change the conclusion regarding the phase transition of H$_2$
  alone, as the species do not remain mixed when H$_2$ alone condenses.} The precipitation of formed ``comets'' and
``planetoids'' in a gravitational field, separating the gaseous and condensed phases, is also a fascinating aspect that
is as yet impossible to capture with traditional hydrodynamical codes, but possible with the molecular dynamics
approach.  We will pursue further simulation work, including more specific astrophysical applications.

\begin{acknowledgements}
  This work is supported by the STARFORM Sinergia Project funded by the Swiss National Science Foundation.  We thank the
  LAMMPS team for providing a powerful open source tool to the scientific community.  We thank the referee for a
  thorough reading of the manuscript and constructive comments, which substantially improved the paper.
\end{acknowledgements}

\bibliographystyle{aa} 
\bibliography{FP2015} 

\end{document}